\documentclass[a4paper,twoside,12pt]{report}
\usepackage{graphics}
\usepackage{latexsym,cmmib57}
\usepackage{amsmath}
\usepackage{amsthm}
\usepackage[psamsfonts]{amssymb,amsfonts,eucal}
\usepackage{epsfig}
\usepackage{color}
\usepackage{multirow}
\usepackage{upgreek}
\usepackage[pdftex]{hyperref}
\usepackage{graphicx}
\usepackage{url}
\usepackage{units}
\usepackage{bigstrut} 
\usepackage{enumitem} 
\usepackage{fancyhdr}
\usepackage{color}
\usepackage[toc,page,titletoc]{appendix}
\usepackage{rotating}
\usepackage{booktabs}
\usepackage{todonotes} 
\usepackage{verbatim} 
\reversemarginpar

\renewcommand{\chaptermark}[1]{\markboth{#1}{#1}}
\renewcommand{\sectionmark}[1]{\markright{\thesection\ #1}}      

\usepackage[font=footnotesize,format=hang,labelfont=bf,textfont=it,skip=6pt]{caption}

\begin{document}
\begin{center}
\centering
\begin{minipage}{3.9cm}
\vspace{-1cm}
\flushleft
	\includegraphics[width=3.9cm]{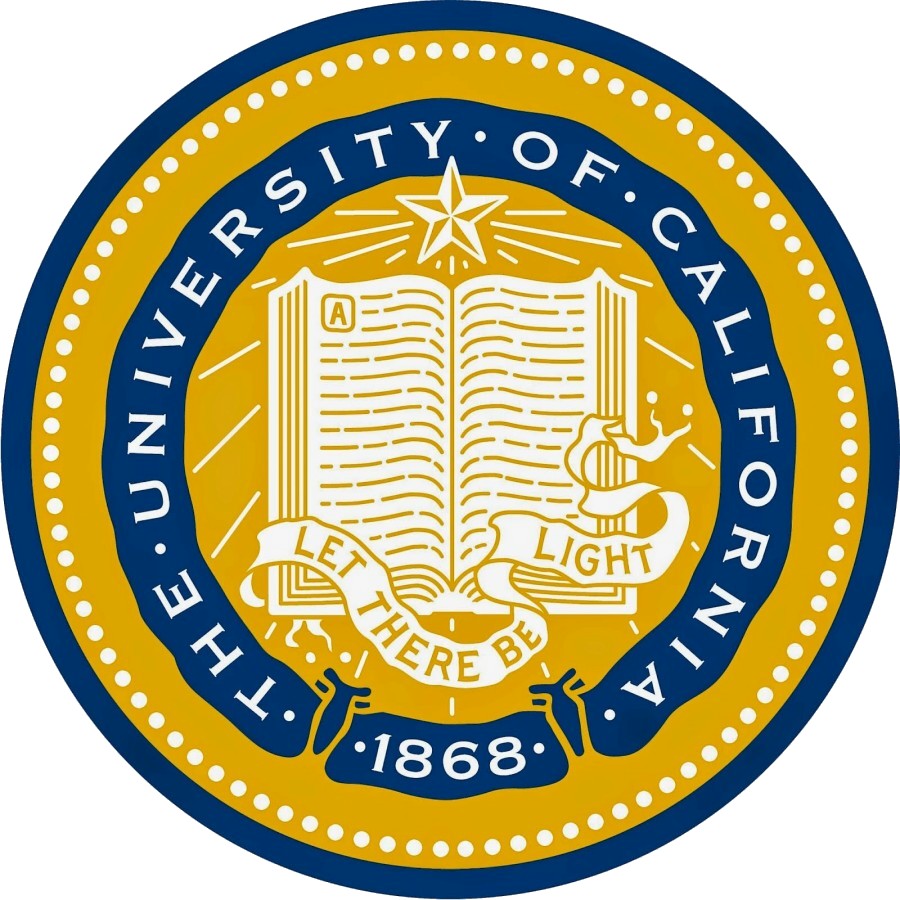}
\end{minipage}
\begin{minipage}{4.9cm}
\vspace{-1cm}
\centering
	\includegraphics[width=4.5cm]{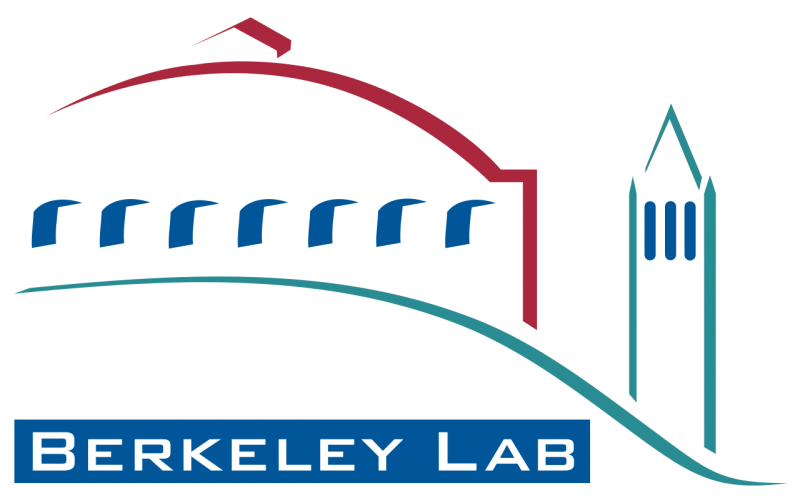}
\end{minipage}
\begin{minipage}{5cm}
\vspace{-1cm}
\flushright
	\includegraphics[width=5cm]{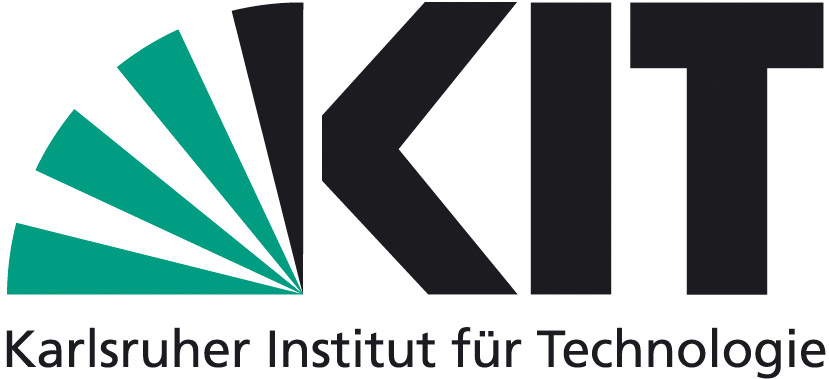}
\end{minipage}

\vspace{1.3cm}
\Large Master of Science in Physics
\vspace{0.8cm}

\huge Investigation of production cross sections,\\
\vspace{0.1cm}
\huge using stacked targets at the 88" Cyclotron\\
\vspace{0.1cm}
\huge with focus on $^{\rm nat}{\rm Fe}(\rm p,x) \phantom{}^{51}{\rm Mn}$ 

\vspace{0.8cm}

\includegraphics[width=420px, keepaspectratio]{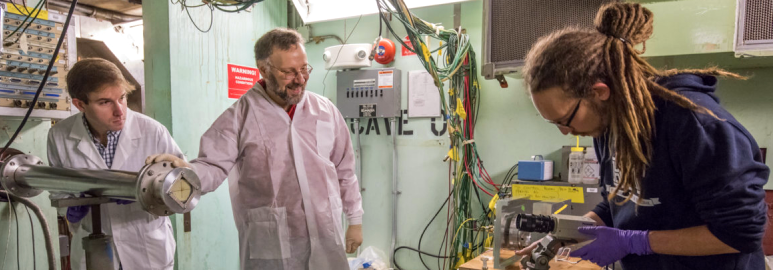}\\
\vspace{0.5cm}
\Large {\it by}\\
\Large {\it Alexander Springer}\\
\ \\
\ \\
\begin{minipage}{4.5cm}
\centering
\large Department of Nuclear Engineering\\
\large University of California Berkeley
\end{minipage}
\begin{minipage}{4.5cm}
\centering
\large Lawrence Berkeley\\
\large National Laboratory
\end{minipage}
\begin{minipage}{4.5cm}
\centering
\large Fakult\"at f\"ur Physik\\
\large Karlsruhe Institute of Technology
\end{minipage}\\
\ \\
\normalsize July, 2017\\
\thispagestyle{empty}
\end{center}
\newpage
\thispagestyle{empty}
\mbox{}

\vfill
\begin{figure}[h!]
\begin{center}
	\includegraphics[width=420px, keepaspectratio]{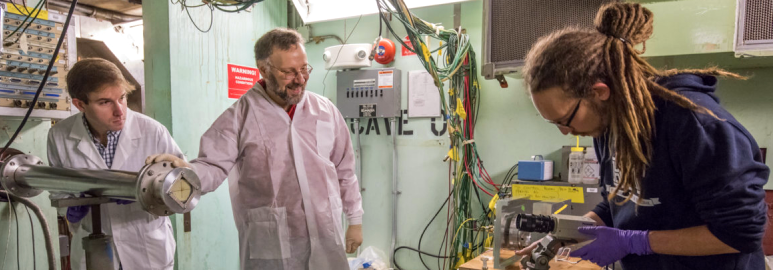}
	\caption{Preparing the phosphorus coated window for tuning the beam prior to the experiment (left to right, Andrew Voyles, Prof. Dr. Lee A. Bernstein, Alexander Springer) \cite{LBLarticle}}
	\label{fig:phosphor}
\end{center}
\end{figure}
\vspace{0.5cm}
\begin{center}
	\textbf{Abstract}
\end{center}

\emph{\textbf{With the investigation of the production cross section of \textsuperscript{51}$\!$Mn off of \textsuperscript{nat}$\!$Fe in mind, two stacked target experiments were conducted at the Berkeley Lab 88-Inch Cyclotron. Proton beams with energies of 25 and 55$\,$Me$\!$V result in a total of 14 data points for various reactions. Gamma spectra of the activated foils were taken and analyzed, then cross sections calculated. }}
\newpage

\pagenumbering{arabic}
\renewcommand{\chaptermark}[1]{\markboth{#1}{#1}}
\renewcommand{\sectionmark}[1]{\markright{\thesection\ #1}}  
\pagestyle{fancy}
\fancyhead[LE,RO]{}
\fancyhead[RE]{\slshape \nouppercase{\leftmark}}
\fancyhead[LO]{\slshape \nouppercase{\rightmark}}
\fancyfoot[LO,LE]{}
\fancyfoot[CO,CE]{\thepage}
\fancyfoot[RO,RE]{}

\tableofcontents
\clearpage
\listoffigures
\listoftables
\clearpage
\leavevmode\thispagestyle{empty}\newpage

\chapter{Introduction}\label{chap:intro}
\thispagestyle{empty}

\section{Nuclear Physics}
To the ancient Greeks the atom was the smallest building block of matter, hence the name from the ancient Greek word \'atomos, meaning indivisibly. The first one to discover the truth was Joseph Thomson who postulated the role of the electron as charge carrier in the atom. In his plum pudding model he assumed there must be some positive charge carrier as well and thought both are distributed evenly throughout the atom. Ernest Rutherford went a step further with his famous experiment, shooting alpha particles onto a thin gold foil. What he found was, that most of the time the $\upalpha$ would just go straight through. Sometimes however the $\upalpha$ would be scattered at different angles, even reflected backwards. Rutherford found the atomic nucleus, the literal core, and concluded it has to be a fraction of the size of the atom but comprises almost all of its mass and is highly charged.\par
It is known now, that the nucleus consists of positively charged protons and neutrons without charge, which are completed by an electron cloud to make up a neutral atom. Charged atoms are called ions. They can have additional electrons which makes them negatively charged, or they could lack electrons making them positively charged.\par
A nucleus is described by the amount of protons $Z$ (atomic number) and neutrons $N$. The mass number $A\!=\!Z\:\!\!+\:\!\!N$ is used in the common notation, where $X$ is a certain element: $\phantom{}_Z^AX$ often times the atomic number gets omitted.

\subsection{Radioactive Decay}
Quantum chromodynamics describes the constituents of protons and neutrons called quarks. Changing one of these quarks allow them to change into each other. This so called beta decay changes nuclei fundamentally producing another element with different properties. The beta minus decay changes a neutron into a proton via the emission of an electron and an electron anti-neutrino with all three products carrying away the mass difference.
\begin{equation}
	\phantom{}_Z^AX \rightarrow \phantom{}^A_{Z+1}X^* + \rm e^- + \bar{\upnu}_e 
\end{equation}
The newly formed nucleus could be left in an excited state, in which case it deexcites via the emission of characteristic photons. By detecting these gamma rays it can be determined what isotope, beta decayed in the first place. 
\begin{figure}[htb]
\centering
	\includegraphics[width=200px, keepaspectratio]{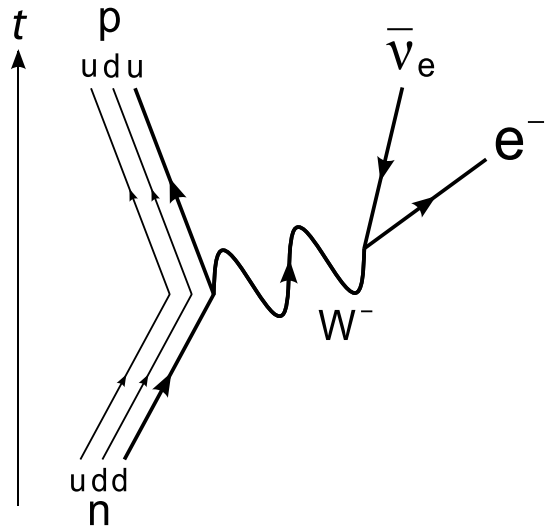}\\
	\caption{Feynman diagram of the beta minus decay. \cite{dec-beta}}
	\label{fig:dec-beta}
\end{figure}\par
The beta plus decay in turn changes a proton into a neutron emitting a positron and an electron neutrino.
\begin{equation}
	\phantom{}_Z^AX \rightarrow \phantom{}^A_{Z-1}X^* + \rm e^+ + \upnu_e  
\end{equation} 
Beta plus decay can only happen within a system that can lend the energy necessary to produce a neutron in the residual nucleus. Unbound protons seem to be stable, or at least they have a half-life exceeding the age of the universe. Free neutrons on the other hand, are able to decay via $\upbeta^-$.\par
Yet another process, results in the transformation of a proton into a neutron, called Electron Capture (EC). One of the inner electrons of the atom is absorbed by the nucleus and performs in the reaction $\rm p+e^-\rightarrow n+\upnu_e$ releasing a monoenergetic electron neutrino. Sometimes these variations of the beta decay are called inverse beta decay, but this term should be reserved for reactions including the neutrino not as product but active participant. The fact that they are part of the same mechanism can be clarified by redrawing the Feynman diagram in Figure \ref{fig:dec-beta}. \par
The only other transmuting decay, of concern here, is alpha decay\footnote{Stated for completeness: Spontaneous fission and beta-delayed proton or neutron emission are also decay processes.}. It is the ejection of a helium nucleus consisting of two protons and two neutrons which are tightly bound in the alpha particle. For heavy nuclei this is a way to get to a more stable state, because it reduces the repulsive Coulomb force. The nuclear force (a residual of the strong force) that keeps the nucleons together in the first place, is powerful but only acts over a short distance. The bigger a nucleus gets the larger the electrostatic repulsion between its constituent protons becomes. With the release of an $\upalpha$, positive charge gets carried away, leaving behind a more stable nucleus.\par
Two more effects are part of radioactive decay. These do not change the atomic number of the nucleus decaying. They are mechanisms to release excess energy on the way back into the ground state. The excitation usually decays in this fashion very quickly ($\approx\!10^{-15}\,\rm s$). Other excited states can last much longer and are then called meta stable states, and the associated nucleus, isomer.\par 
The first process is known as Internal Conversion (IC), where electrons can be emitted. These electrons come from the electron shell, not the nucleus itself, which differentiates it from beta decay. Necessary for IC to work is the fact that the electron wave function overlaps with the atomic nucleus to some degree. The energy to eject these inner electrons (most common K-shell or 's' states from other shells) comes directly from the nucleus. Afterwards a cascade of x-ray photons and Auger electrons is emitted by electrons shifting down to fill the vacancy. IC is possible whenever gamma decay is possible, unless the atom is fully ionized. It dominates for transitions with energies near an atomic ionization threshold.
The second is gamma decay, the more common and fast process of nuclear deexcitation. This is the strongest penetrating radiation of the ones discussed so far. As hinted before, this decay succeeds other forms of decay, like alpha and beta decay. Plain and simple, the gamma rays can carry away varying amounts of energy from the nucleus. The amount of energy is given by the difference between the two energy levels involved. This allows for the study of these energy levels and therefore nuclear structure in general.

\subsection{Decay law}
The radioactive decay discussed so far, is a passive process from the observers point of view. Radioactive decay is random, so much so, that it is often used in thought experiments like Schr\"odinger's cat as the random element. Independent from the type of decay is its statistical nature. It is impossible to predict when a specific nucleus will decay, but the decay process can be quantified for a multitude of nuclei $N$, using probability. Different isotopes may decay with different rates $\lambda$, but following the same exponential decay law, in which the initial amount $N_0$ gets reduced by ${\rm d}N$.
\begin{equation}\label{eq:decay}
	\frac{{\rm d}N}{{\rm d}t}=-\lambda N
\end{equation}
If there is only one isotope decaying into a stable one and no additional production contributing, the solution to the equation above is:
\begin{equation}\label{eq:decay2}
	N(t)=N_0 e^{-\lambda t}
\end{equation}
\begin{figure}[htb]
\centering
	\includegraphics[width=350px, keepaspectratio]{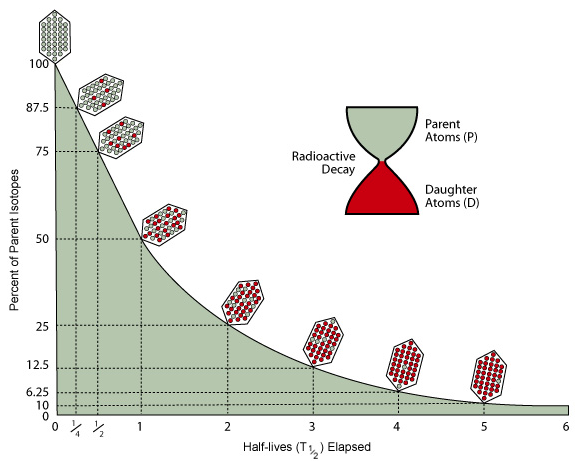}\\
	\caption{Ilustration of the exponential nature of radioactiv decay \cite{dec-law}}
	\label{fig:dec-law}
\end{figure}\par
Shown in the visual representation of the decay law in Figure \ref{fig:dec-law} is the half-life $t_{\nicefrac{1}{2}}$ on the x-axis. After one elapsed half-life, half of the initial number of nuclei will have decayed.
\begin{equation}
	t_{\nicefrac{1}{2}}=\frac{\ln(2)}{\lambda}
\end{equation}
The isotope a radioactive material decays into is called the daughter. One step further down in the decay chain and the term granddaughter is used.\par
If the substance in question is not geological in nature, but rather being simultaneously formed as part of a nuclear physics experiment, both production and decay have to be taken into account. This can be achieved by including a production rate $R$ into Equation \ref{eq:decay}:
\begin{equation}
	\frac{{\rm d}N}{{\rm d}t}=R-\lambda N
\end{equation}
For a time $t_{\rm i}$ production and decay are considered, after that, for the time $t_{\rm cool}$ only decay proceeds. The solution for this scenario looks like:
\begin{equation}\label{eq:decay-n}
	N(t)=\frac{R}{\lambda}(1-e^{-\lambda t_{\rm i}})e^{-\lambda t_{\rm cool}}
\end{equation}
This equation can gain complexity quickly when production and decay of multiple decays are taken into account.

\subsection{Nuclear Reactions}\label{sec:nuc-reac}
If radioactive decay is passive, then nuclear reactions are active. A nuclear reaction is a process set in motion by some event, it is not happening randomly. The event to occur usually involves a particle getting close enough to the nucleus to react. Unless there is a nuclear chain reaction possible, particles have to keep coming in to sustain the reaction. In experiments this kind of beam is supplied by a reactor or an accelerator.
Many different reactions are possible but often the simplest ones, involving the fewest number and type of particles, are the most probable. All nuclear reactions conserve energy. With this in mind the reaction Q value is a way to quantify how probable a reaction is. The reaction Q value is defined as the initial minus the final mass energy, which is nothing else than the residual excess kinetic energy ($T$) of a reaction.
\begin{align}
	Q &= (m_{\rm initial} - m_{\rm final})c^2\\
	Q &= T_{\rm final} - T_{\rm initial}
\end{align} 
If its positive, and the higher the value, the more probable it is for the reaction to occur. Falling below zero and there has to be additional energy put into the system, otherwise the reaction will not take place.\par
In this work, the production of manganese via nuclear reaction is in the foreground. One of the reactions is the following:
\begin{equation}\label{eq:react}
	\phantom{}^{54}{\rm Fe} + \phantom{}^1{\rm H} = \phantom{}^{51}{\rm Mn} + \upalpha \underbrace{-3146.57(\pm 0.58)\,{\rm keV}}_\textrm{Q value}
\end{equation}\cite{nndc-q}
A hydrogen nucleus or simply a proton, will react with a $^{54}\rm{Fe}$ nucleus to form $^{51}\rm{Mn}$ and an $\upalpha$. In this form the proton has no initial energy and the Q value for the reaction is negative. Only when the proton is given enough initial energy the reaction will start to occur. Testing this reaction (but more broadly, over all natural abundant iron isotopes) at different beam energies and determining how often the reaction occurred is a basic, but good, description of the experiment in this work.\par
For a more compact and more clear syntax, the nuclear reaction notation is used. The target is followed by the projectile and ejectile in parentheses and the residual nucleus. Expressing Reaction \ref{eq:react} in this way, turns it into: $\rm ^{54}Fe(p,\upalpha) \phantom{}^{51}Mn$.\par
The nuclear reaction cross section $\sigma$ is a measure of the probability that a given reaction occurs. The cross section is defined as an area per nucleus with the SI unit $\rm m^2$. Because the cross section is usually very small, the barn or millibarn $\rm 1\, mb = 10^{-31}\, m^2$ is commonly used. This value can be bigger or smaller than the geometrical surface area of the target nucleus visible to the beam. For an average nucleus with a radius of $6\,\rm fm$ the geometric area would be about one barn $1\,\rm b = 10^{-28}\, m^2$.
\begin{equation}\label{eq:cs}
	\sigma = \frac{R}{\Phi \cdot N_{\rm T}}
\end{equation}
Equation \ref{eq:cs} defines the reaction cross section with $R$ being the reaction rate (events per second), $\Phi$ the incomming particle flux (particles per second) and $N_{\rm T}$ the number of target nuclei per area. \cite{krane}

\section{Gamma spectroscopy}
Radioactive materials emit photons which, when detected, can tell us what is happening in the material. These photons are usually not part of the visible spectrum but on the keV-MeV scale and are defined as gamma rays, if they come from the nucleus and not an atomic process. Most $\upgamma$-ray detectors use a scintillation crystal like sodium iodide (NaI) or a semiconductor. A high-purity germanium detector (HPGe) is a good example of the latter. When ionizing radiation hits the biased semiconductor it deposits energy that can be measured as current. A closer look reveals that the electrons move from the valence into the conduction band, given enough energy, where the electric field moves them along. In the same manner the electron gap or 'hole' moves in the opposite direction. Both currents contribute to the signal.\par
Since the gap between the two bands is quite narrow for germanium, it is very precise in its reaction. The downside is, its susceptibility to thermal excitation. For that reason, a HPGe detector has to be cooled.\par 
The acquisition relies on fast electronics and a computer, running software to record the gamma ray spectra and save them to disk.

\chapter{Motivation}
Manganese is a promising candidate for medical applications. It has been used in magnetic resonance imaging (MRI) as contrast agent, but the doses involved are toxic to humans. Manganese is better suited for use in positron emission tomography (PET), yielding better contrast. $^{52\rm g}\rm{Mn}$ is promising but far from perfect because of its long half-life ($5.59\,\rm days$) and high-energy gamma emissions. Nevertheless, various studies are being conducted and there are uses like ImmunoPET for which production of $^{52}\rm Mn$ is explored \cite{immunopet}. Far better suited in this regard is $^{51}\rm{Mn}$ with a short $46.2\,\rm{min}$ half-life\footnote{Graves et al., determined with high precision the half-life to be $45.6\,\rm{min}$\cite{graves-half-life}. The difference is not noticeable in this work, and since these findings are yet to be published, the Nudat value was used \cite{nndc-nudat}.} and no prominent gamma emissions. The only considerable unwanted dose is coming from the daughter $^{51}\rm{Cr}$ with a half-life of $27.7\,\rm{days}$, which emits a $320\,\rm{keV}$ $\upgamma$-ray for every 10th decay.\par
Since $^{51}\rm Mn$ is a promising candidate, it is of considerable interest to investigate how feasible the production is, as well as its medical value. Production channels of all sorts are being looked at and evaluated, even though there has not been a lot of experiments regarding low energy accelerators. Daube et al. and Klein et al. produced $^{51}\rm{Mn}$ by the $^{50}\rm Cr(d,n)$ reaction with Klein using a pure metal powder target and Daube an oxid powder \cite{daube} \cite{klein1} \cite{klein2}. None of the two managed to fabricate a robust target exceeding $4\,\rm{\upmu A}$ in irradiation current handled \cite{hichwa}. Much more is needed though to create a usable amount.\par 
One criterion is the abundance of the target material used as base to produce a certain isotope. With the experiment conducted in this work, a very abundant and easy to come by target is used: natural abundant iron. While the yield might be higher and purer with using enriched targets, e.g. iron enriched in $^{54}\rm{Fe}$, those targets are also more expensive to produce. Graves et al., did not only utilize enriched targets (own production), but also performed in vivo testing in mice \cite{graves51Mn}. The goals being to develop the production via $^{54}\rm{Fe}(p,\upalpha)$ and characterization of the in vivo behavior of $^{51}\rm{MnCl}_2$ in mice as well as overall dose. Graves et al., argues that manganese, because of its role in mammalian biology \cite{mambio} makes a good tracer and that $^{51}\rm{Mn}$ out of all isotopes of manganese is best suited for PET.\par
\begin{figure}[t]
\centering
	\includegraphics[width=300px, keepaspectratio]{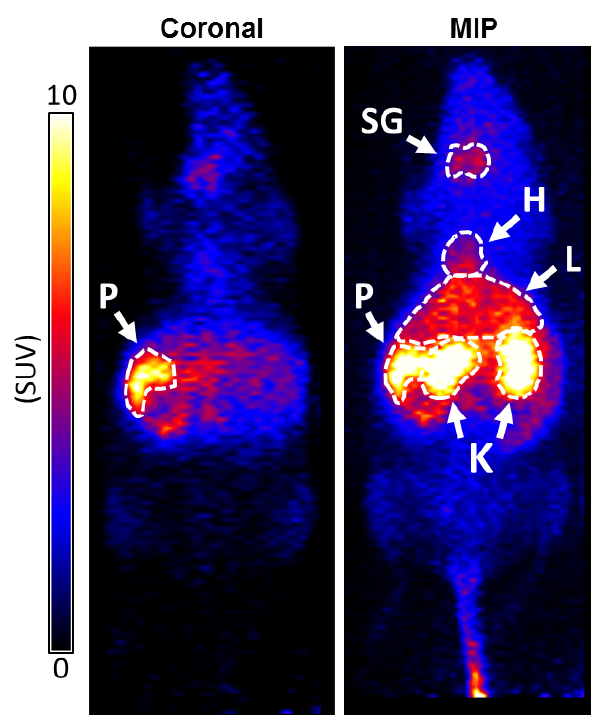}\\
	\caption[Mouse, one hour after being injected with the $^{51}\rm Mn$ tracer, Coronal slice and maximum intensity projection. Highlighted areas: salivary gland (SG), heart (H), liver (L), pancreas (P) and kidneys (K). \cite{graves51Mn}]{Mouse, one hour after being injected with the $\mathit{^{51}\!Mn}$ tracer, Coronal slice and maximum intensity projection. Highlighted areas: salivary gland (SG), heart (H), liver (L), pancreas (P) and kidneys (K). \cite{graves51Mn}}
	\label{fig:mouse}
\end{figure}

In their study Graves et al., have shown that it is possible to produce and utilize $^{51}\rm{Mn}$ from enriched $^{54}\rm{Fe}$. Furthermore, they have discussed the usefulness for medical applications. From the distribution and dose in mice as visible in Figure \ref{fig:mouse} and dose measurements ex vivo, they determined that PET imaging of pancreatic beta cells with $^{51}\rm{MnCl}_2$ is promising. Further studies are needed to quantify the benefits of the variety of PET applications conceivable.\par
\par
In the end the production from natural iron will produce contaminates alongside with the desired isotope due to reactions from the heavier stable isotopes of iron. Whether this is a plausible production route, depends on how much effort it is to clean the product, a question that remains to be answered. Time is an important factor here, the separation of unwanted isotopes has to be happening quickly, otherwise most of the main isotope will have decayed away. Another factor is recoverability of the target. How much if any of the used target can be reused?\par
The focus of this work lies in measuring the production cross-section as function of the beam energy of $^{\rm{nat}}\rm{Fe}(p,x) \phantom{}^{51}\rm{Mn}$. Underneath a certain threshold there are no contributions from the other iron isotopes expected, making this effectively a measurement of the $^{54}\rm{Fe}(p,\upalpha) \phantom{}^{51}\rm{Mn}$ cross section. Knowing the production cross section is a necessity to design an efficient production path. The beam energy for production will be tuned to the peak of the cross section, unless other factors come to play, like additional contamination channels opening up.\par
There are ways to pinpoint the cross section from data and simulations already existing. How these are holding up and compare to the findings in this work, is being discussed in Chapter \ref{chap:conc}.
\newpage
\leavevmode\thispagestyle{empty}\newpage

\chapter{Experimental Setup}
The approach used in this experiment features a charged particle beam incident on a stack of target foils.
This approach allows for the simultaneous measurement of the reaction cross section at different energies. This is achieved by alternating target and monitor foils with well-known proton-induced reaction cross sections, with thicker pieces of aluminum that degrade the beam, called degraders.\par
Two stacks were run at two different initial beam energies: $25\,\rm{MeV}$ and $55\,\rm{MeV}$ protons. The experimental work was conducted at the 88-Inch K-140 cyclotron at Lawrence Berkeley National Laboratory (LBNL) in Berkeley, California. Irradiation took place on the 15th and 16th of December 2016.

\section{Preparations}

\subsection{Probed Reaction}\label{chap:react}
In this work the focus lies on the production of $^{51}\rm{Mn}$ off natural iron. The only reaction contributing to the production at low energies is 
\begin{equation}
	 ^{54}\rm{Fe}(p,\upalpha) \phantom{}^{51}\rm{Mn}
\end{equation}
For higher energies more channels are opening up, especially the much more abundant $^{56}\rm{Fe}$ contributes (see Table \ref{tab:abun}). 

\begin{table}[htb]
	\centering
	\caption{Composition of natural abundant iron \cite{nndc-nudat}}
	\begin{tabular}{cr}
		\hline
		Isotope & Abundance ($\%$) \bigstrut[t]\\
		\hline
		$^{54}\rm{Fe}$ & 5.845 $\pm$ 0.035 \bigstrut[t] \\
		$^{56}\rm{Fe}$ & 91.754 $\pm$ 0.036\\
		$^{57}\rm{Fe}$ & 2.119 $\pm$ 0.010\\
		$^{58}\rm{Fe}$ & 0.282 $\pm$ 0.004\\
		\hline
	\end{tabular}
	\label{tab:abun}
\end{table}

Q values of the different reactions indicate the threshold energies required to open a given exit channel (more in Chapter \ref{sec:nuc-reac}). The National Nuclear Data Center (NNDC) supplies a tool to calculate Q values easily one by one (see \cite{nndc-q}). However, since the Q value only provides information about reaction thresholds, it is advisable to use an extensive reaction code like Talys \cite{talys}. For a given projectile, energy and target material, Talys will utilize state of the art nuclear models to simulate the nuclear reaction. It provides which isotopes are likely to be produced, what the cross sections are as well as other observables.\cite{talys} \par
Following irradiation the different radioactive nuclides produced will beta-decay into excited nuclear states, producing $\upgamma$-rays that can be used to quantify the yield of a specific reaction channel. Some isotopes actually have the same energy levels which will result in the exact same $\upgamma$-rays emitted. In other cases the line that needs to be observed will be drowned out by a very high nearby activity. To avoid this from becoming apparent after the fact, a thorough investigation of the decay radiation is in order. A trusted source for this data is the National Nuclear Data Center in Brookhaven National Laboratory, which maintains the Nudat database \cite{nndc-nudat}. For every radioactive isotope, for which there is data found in the system, it lists the decay radiation, including gamma and x-ray. Regarding the intensities, energy and target abundance for every possible residual product, conflicts can be determined and addressed. If the high intensity channels are not conclusive, a lower intensity line might be without contamination or it might be worth looking into the granddaughter as a gauge of a reaction.\par 
$^{51}\rm Mn$ almost exclusively emits $511\,\rm{keV}$ $\upgamma$-rays from the $\upbeta^+$ decay. The positron emitted will find an electron quickly and annihilate resulting in two $511\,\rm{keV}$ $\upgamma$-rays bearing the electrons rest mass. 511s are special because they can come from a lot of different sources which makes it hard to associate them to one particular isotope decaying. Not only do they appear in most decays, but also manifest when a higher energy $\upgamma$ (above $1022\,\rm{keV}$) produces an $\rm e^- e^+$ pair. The positron annihilates, but often only one of the resulting $\upgamma$-rays gets detected.
There are ways to investigate the origin of $511\,\rm keV$ $\upgamma$-rays (see Section \ref{sec:get}), but whether the process will be conclusive or not is hard to say in advance. Luckily there is a second path to explore. $^{51}\rm Mn$ decays into $^{51}\rm Cr$ which itself is also radioactive. This granddaughter of the reaction that is investigated, exhibits a much longer half-life than the Manganese. When it decays, there is a $9.9\,\%$ chance it will emit a $320\,\rm{keV}$ $\upgamma$-ray. None of the other residual products seem to contaminate this line. This will be the main method to quantify the production of $^{51}\rm Mn$.\par
\clearpage
\newpage

\subsection{Stack design}
The target stacks are carefully designed prior to the experiment. Apart from the obvious natural iron foils that are needed for the reaction as just discussed above, there is a number of other components. Tabel \ref{tab:lowstack} lists all the components, of the low energy stack, with their thickness and areal density. The areal density is stated as determined in the characterization process and used later in the analysis. In the Appendix \ref{app:highstack}, Figure \ref{tab:highstack}, the high energy stack design is listed in detail.\par
Foils are cut into square shape, with an approximate edge length of $25\,\rm mm$. Precise geometric measurements are taken and the foils wiped with rubbing alcohol to remove any dirt. The foils are then weight, before they are embedded in kapton tape\footnote{High-Temperature Kapton Polyimide Masking Tape supplied by McMaster-Carr USA \cite{kapton}} and centered on aluminum frames. This is to shield them from damage or pollutants as well as seal the foils off.\par
For each iron foil\footnote{$99.5\,\%$ pure natural iron foils supplied by Goodfellow Cambridge Limited Huntingdon England (\url{www.goodfellow.com})}, certain monitor foils are added. Monitor foils are chosen in regard to the energy region that is studied. The monitor reaction in this region should show a clear shape, best of all cover a peak in the cross section.
These monitors help to figure out later what was actually going on in the stack. They help to determine energy as well as proton flux throughout the stack.
Copper\footnote{Mostly $99.95\,\%$ pure natural copper foils supplied by Goodfellow. Cu14-16: $>99.99\,\%$ pure natural copper from Alfa-Aesar (\url{www.alfa.com}) or Sigma Aldrich (\url{www.sigmaaldrich.com})} and titanium\footnote{Mostly $>99.6\,\%$ pure natural titanium foils supplied by Goodfellow. Ti14-16: $>99.99\,\%$ pure natural titanium from Alfa-Aesar or Sigma Aldrich} were chosen as monitor foils and these monitor reactions with it: 
\begin{align}
^{\rm{nat}}\rm{Cu}(p,x)&^{62}\rm{Zn}\\
^{\rm{nat}}\rm{Cu}(p,x)&^{63}\rm{Zn}\\
^{\rm{nat}}\rm{Cu}(p,x)&^{65}\rm{Zn}\\
^{\rm{nat}}\rm{Ti}(p,x)&^{48}\rm{V}
\end{align}\par
\begin{figure}[h!]
\centering
	\includegraphics[width=435px, keepaspectratio]{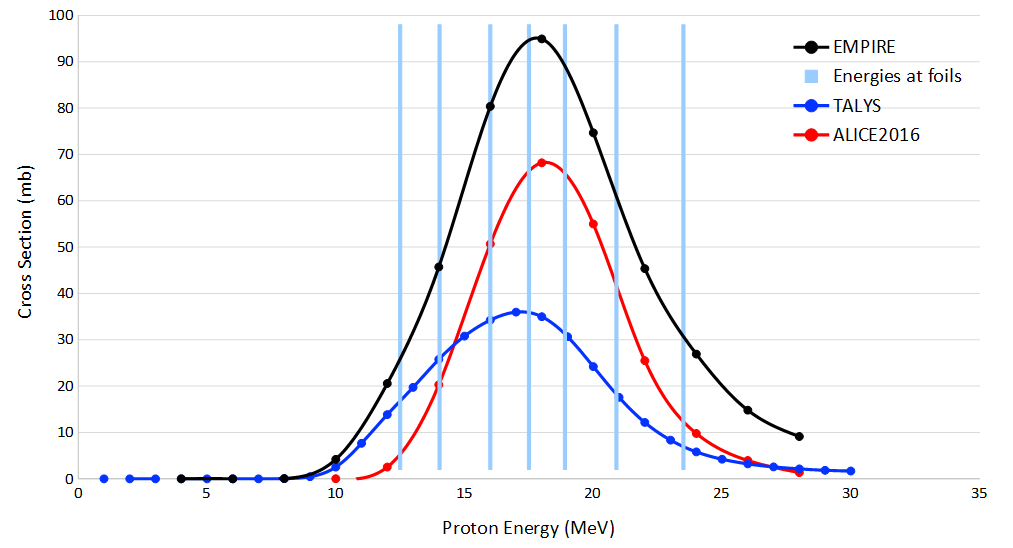}\\
	\caption{Comparison with reaction codes to place foil energies.}
	\label{fig:talys}
\end{figure}

Mentioned earlier, there are also aluminum degraders, placed to space out the energy points. In this process, a rough estimate of how the beam will be degraded down and shift its energy while traversing the stack, is calculated. In Figure \ref{fig:talys} it is probed, where on the cross sections, simulated via different reaction codes, the experiment will yield data points.
This way, the main iron foils can be placed evenly or at energies of special interest. For example around the estimated peak of the cross section to be studied. Again it is helpful to look at the Talys simulation for an estimate. Despite the usefulness of these simulations, one must not forget that the physical reality very well may look different.\par
\newpage
Lastly, there is a stainless steel plate in the front and the back of the stack that act as beam profile monitors.
After the irradiation, Gafchromic film \footnote{Gafchromic EBT3 (\url{www.gafchromic.com}) supplied by Radiation Products Design Inc. Albertville MN (\url{www.rpdinc.com})} will be placed on these plates and left there. The area, where the beam hit the plate will expose the film and show how the beam looked like entering and exiting the stack.
It is actually the emitted beta electrons that develop the film (any ionizing radiation will do that). This film is also used directly in medical treatment to ensure photon beams have the desired shape, which is very similar to the second use of these films in this experiment. That is, to determine if the beam is on target prior to the irradiation (see Chapter \ref{chap:irr}).
\begin{table}[h]
	\centering
	\caption[$25\,\rm MeV$ stack design. Kapton adhesive tape envelopes every thin foil, and additional to $25.4\,\upmu\rm m$ of kapton there is also $38.1\,\upmu\rm m$ of acrylic adhesive slowing down the beam.]{$\mathit{25\, Me\!V}$ stack design. Kapton adhesive tape envelopes every thin foil, and additional to $\mathit{25.4\mu m}$ of kapton there is also $\mathit{38.1\mu m}$ of acrylic adhesive slowing down the beam.}
	\begin{tabular}{cccc}
		\hline
		Foil & Material & \begin{tabular}{c}Apparent \bigstrut[t]\\ thickness ($\upmu\rm m$)\\ \end{tabular} & \begin{tabular}{c}Actual  \bigstrut[t]\\ areal density ($\nicefrac{\rm{mg}}{\rm{cm}^2}$) \end{tabular}\\
		\hline
		SS5 & Stainless steel & 125 & 100.57 \bigstrut[t] \\
		Fe 1 & Iron & 25 & 19.69 \\
		Ti 14 & Titanium & 25 & 10.87 \\
		Cu 14 & Copper & 25 & 17.49 \\
		E9 & Aluminum & 254 & 68.18 \\
		Fe 2 & Iron & 25 & 19.90 \\
		Ti 15 & Titanium & 25 & 10.97 \\
		Cu 15 & Copper & 25 & 17.63 \\
		H1 & Aluminum & 127 & $\sim\,$34.3 \\
		Fe 3 & Iron & 25 & 19.84 \\
		Ti 16 & Titanium & 25 & 10.96 \\
		Cu 16 & Copper & 25 & 17.22 \\
		Fe 4 & Iron & 25 & 19.96 \\
		Ti 17 & Titanium & 25 & 10.88 \\
		Cu 17 & Copper & 25 & 21.91 \\
		Fe 5 & Iron & 25 & 20.03 \\
		Ti 18 & Titanium & 25 & 11.00 \\
		Cu 18 & Copper & 25 & 22.33 \\
		Fe 6 & Iron & 25 & 20.05 \\
		Ti 19 & Titanium & 25 & 11.01 \\
		Cu 19 & Copper & 25 & 22.32 \\
		Fe 7 & Iron & 25 & 20.11 \\
		Ti 20 & Titanium & 25 & 11.06 \\
		Cu 20 & Copper & 25 & 22.34 \\
		SS6 & Stainless steel & 125 & 100.99 \\
		\hline
	\end{tabular}
	\label{tab:lowstack}
\end{table}\par
\clearpage
\newpage

\section{Equipment}
The beam for this activation experiment was supplied by the 88-Inch Cyclotron at LBNL. This is a $K\!\!=\!140$ machine which started operations in 1961. It is worth mentioning that the inventor of the cyclotron not only did so at the University of California Berkeley but also founded the LBNL, at that point called UC Radiation Laboratory. Ernest Orlando Lawrence filed his patent in 1932 and built the first cyclotron which is on display at CERN, Geneva. \cite{hp-cyc}
Some others were coming up with the same idea around that time, but no one pursued it for various reasons. Lawrence received the Nobel Prize in physics for his invention in 1939.

\subsection{Cyclotron}\label{sec:cyc}
\begin{figure}[htb]
\centering
	\includegraphics[width=420px, keepaspectratio]{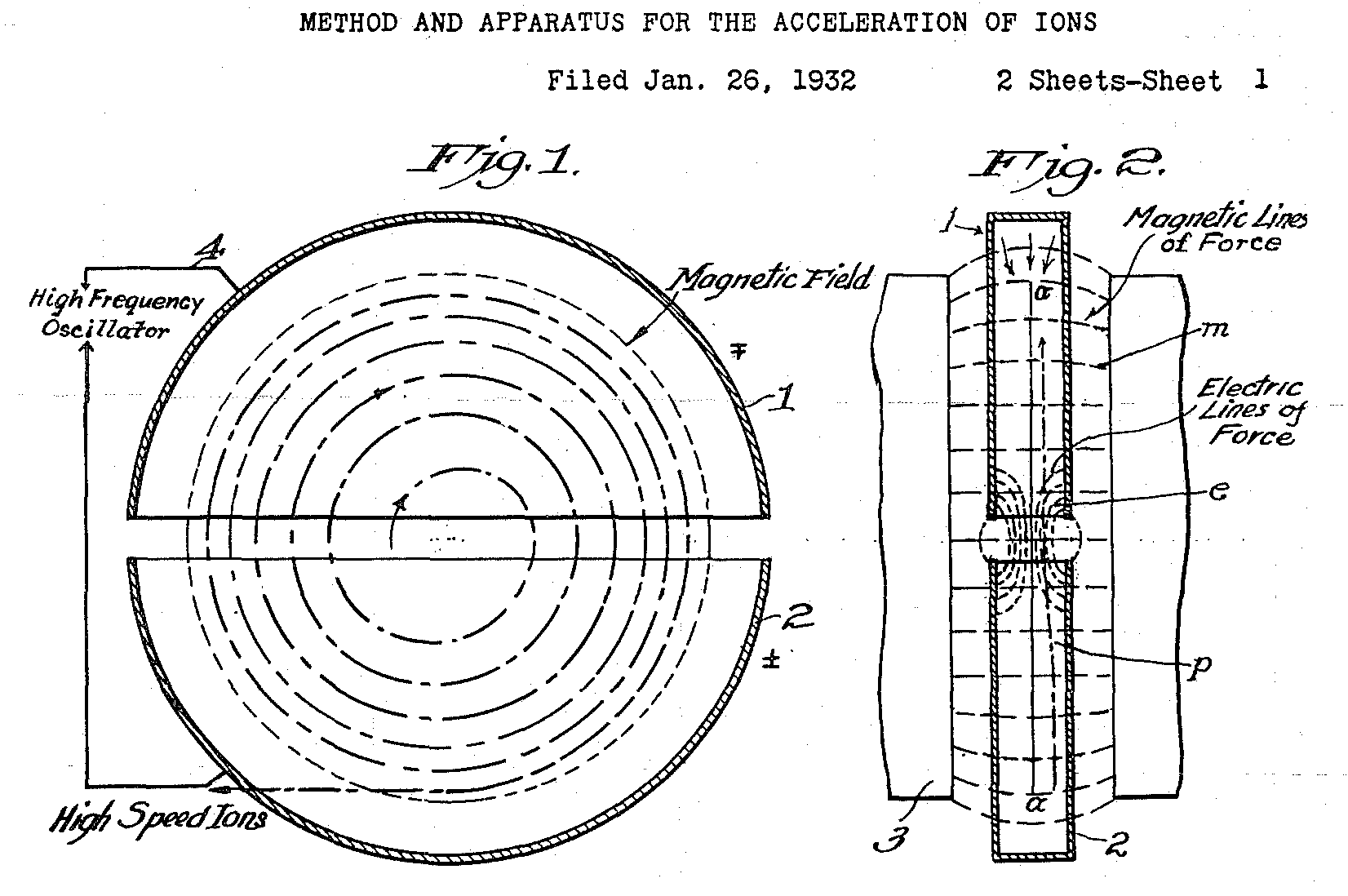}\\
	\caption{Schematic view from the original patent by E. Lawrence \cite{pat-cyc}.}
	\label{fig:cyc-pat}
\end{figure}

Generally a cyclotron accelerates particles in a spiral, with the centripetal force supplied by a static perpendicular magnetic field. The accelerating force comes from an electric field ($\vec{E}$) in the gap between the electrodes, called "Dees" because of their original shape. Two of them are visible in the original illustration in \ref{fig:cyc-pat}, one on top and one on the bottom. The voltage alternates with a high frequency, so that the beam is never being decelerated. This results in the force $\vec{F}=q\cdot \vec{E}$ pushing and pulling in rapid succession on the charge $q$ of the particle. The radius of the particles increases with energy each round and the particles will be extracted from the outside at a certain radius and therefore exit energy. 
\par
The magnetic field $B_{\rm z}$ and the angular frequency $\omega_{\rm z}$ have to be tuned for the cyclotron to work properly. Both are described in regard to the z-axis which the particles revolve around.
\begin{align}
	&\qquad &\qquad  \omega_{\rm z}=\frac{qB_{\rm z}}{m}\\
	&\qquad &\qquad  && q &:\ \text{particle charge}\nonumber\\
	&\qquad &\qquad  && m &:\ \text{particle mass}\nonumber
\end{align}
In order to achieve higher beam energies, it is important to account for relativistic behavior of the particles that make up the beam. To keep the angular frequency constant, at higher energies, one way to go is, to shape the magnetic field accordingly. The field strength has to increase with energy and therefore radius. This machine is referred to as isochronous cyclotron, since it is different from a synchrocyclotron in that the magnetic field is not changing with time. For this to work, the angular frequency has to be constant, following this relation \cite{iso-cyc}:
\begin{equation}\label{eq:cyc}
	\omega_{\rm z} = \frac{q B_{\rm z}(r(E))}{m(E)} = \rm const.
\end{equation}
The 88-Inch Cyclotron is an isochronous cyclotron. A small relativistic correction has to be made to the mass in Equation \ref{eq:cyc} exceeding a certain energy. This is generally the case for all accelerators with $K>35$. \par
Every volume, the beam moves through, is evacuated with vacuum pumps running constantly to keep it that way. In between there are steel pipes keeping the atmosphere out. Dipole magnets are used to change the direction of the beam. Quadrupole magnets focus the beam. Two are used in succession, and rotated $90^\circ$ to each other, since they focus in one direction but act defocusing the other way.\par
\begin{figure}[p]
\centering
	\includegraphics[width=420px, keepaspectratio]{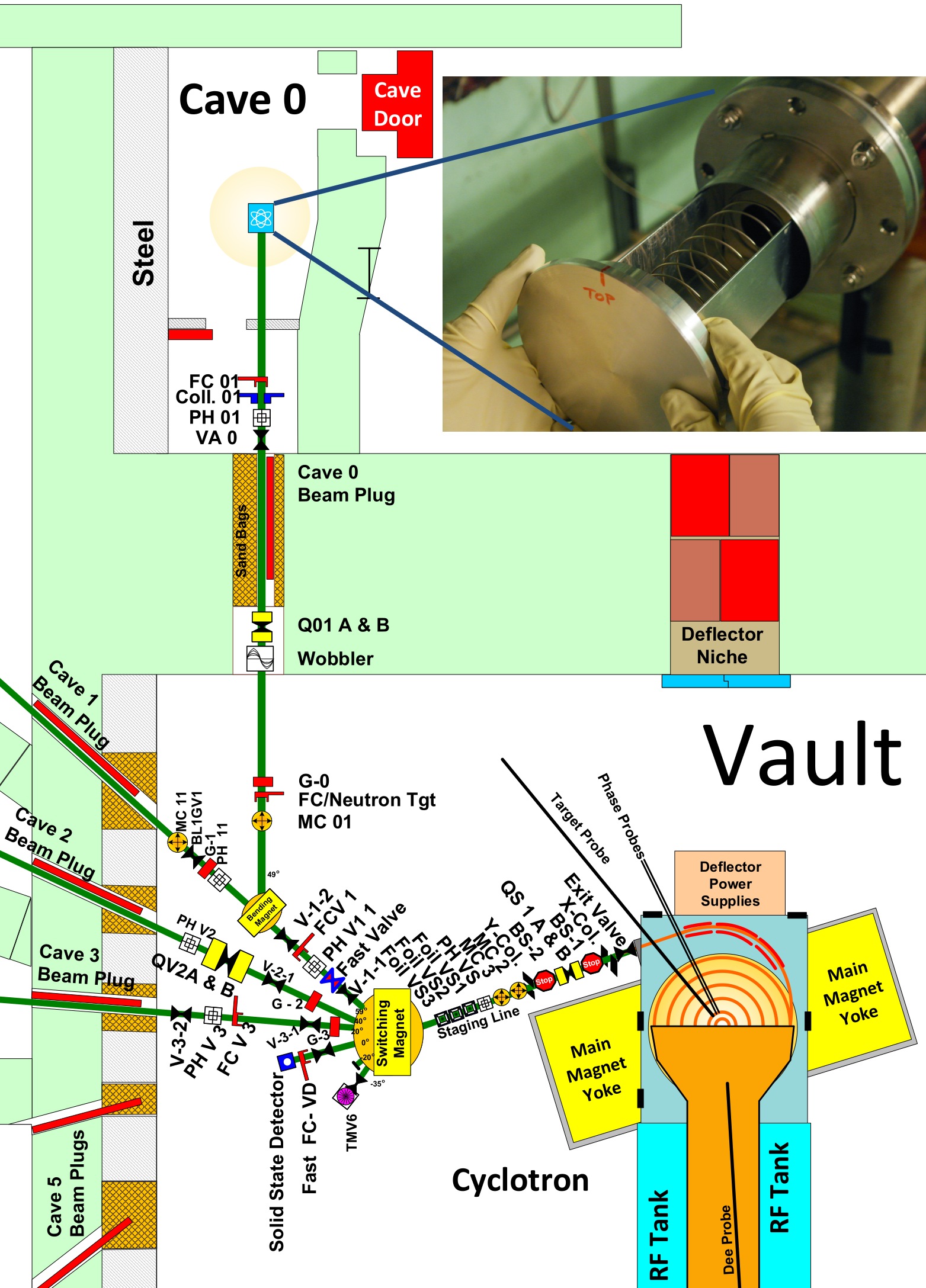}\\
	\caption{88" Cyclotron and beam path to Cave 0}
	\label{fig:cyc-path}
\end{figure}

The experiment is conducted in Cave 0, which is heavily shielded and allows for the highest prompt radiation levels. After the proton beam is extracted from its cyclotron orbit, it leaves the magnetic field and proceeds toward the switching magnet, a dipole. Collimators shave the sides of the beam off along two axes and two quadrupoles are in use. See Figure \ref{fig:cyc-path} to follow the beam path described. The switching magnet steers the beam in the different Caves 0/1, 2, 3 and potentially 5. A second bending magnet turns the beam toward Cave 0. The beam is led through the shielding vault wall into Cave 0, where it hits the targets in the target holder at the end of the beam pipe. 
\par
\clearpage
\newpage

\subsection{Detection}
Different from other particle physics or nuclear physics experiments, there are no detectors at play during the experiment. Instead, the activated targets are extracted from the beam line after the experiment and brought over to the counting lab. The counting lab is equipped with a high-purity germanium (HPGe) detector. Encased in granite and lead to reduce background, the detector points upwards towards its sliding hatch. Samples can be placed at various distances from the face of the detector on an acrylic rack. This is necessary to accommodate for different strengths of activation. If the sample activity level is above the level that the HPGe detector can tolerate (typically, $\approx\! 1\,\rm{kHz}$) it is placed further away from the detector to reduce dead time. Dead time is the time after a gamma ray is detected, and the detector is blind for others. If the rate is high, there will be a lot of counts that get lost in that blind spot. It is favorable to keep the dead time under $10\,\%$. The software in use called Maestro \cite{maestro}, helps to quickly determine, if a sample should be placed at a different position, since it shows live and dead time as well as the percentage.\par
\begin{figure}[hb]
\centering
	\includegraphics[width=180px, keepaspectratio]{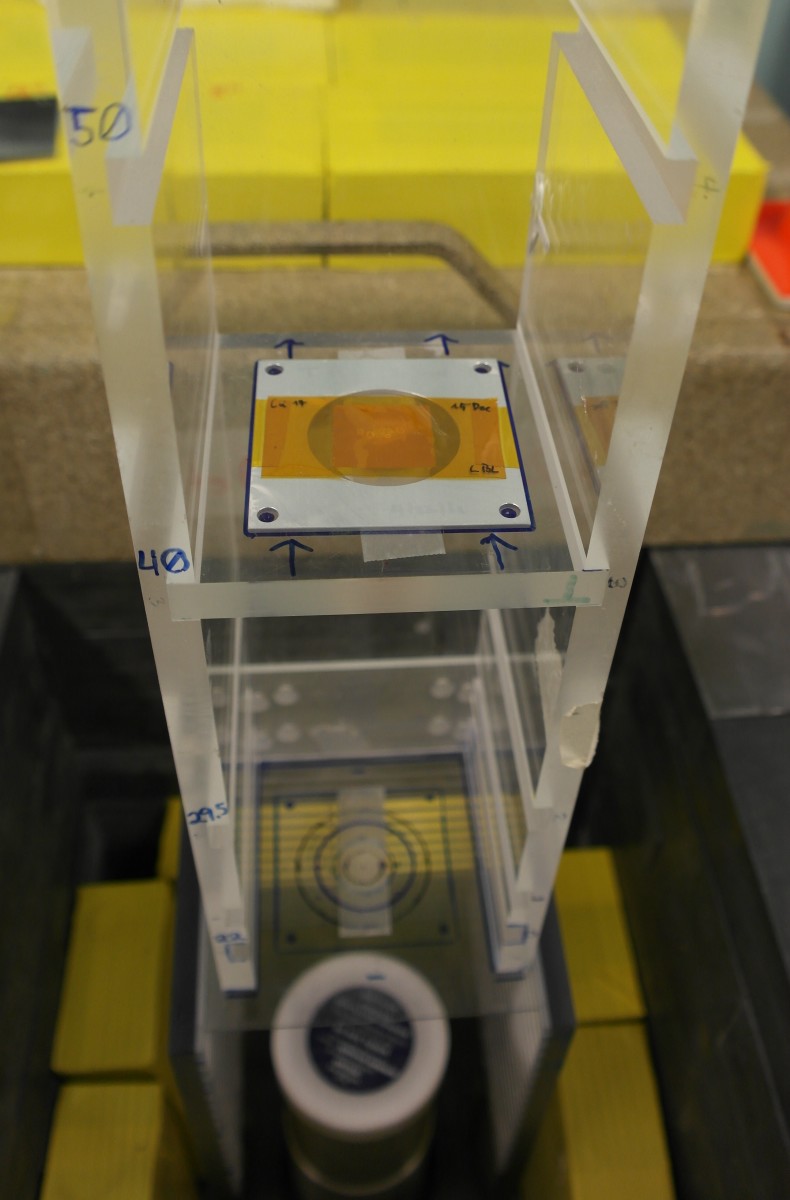}\\
	\caption{Foil in place for counting.}
	\label{fig:det}
\end{figure}

The foils can be placed anywhere from $1\,\rm{cm}$ to $18\,\rm{cm}$. After that, an extension is used (see Figure \ref{fig:det}), to allow the sample to be placed $22$, $29.5$, $40$, $50$ and $60\,\rm{cm}$ from the detector.
The germanium crystal has to be cooled and gets fed with liquid nitrogen from a Dewar (vacuum flask). To protect the crystal, there is a cap in place, with a thin $0.5\,\rm{mm}$ window made from beryllium, which leaves a $4\,\rm{mm}$ gap to the crystal.\par
\par
The distance from the detector can also greatly effect the usefulness of the recorded spectrum. Counting close to the detector can result in counts for specific peaks being lost. This happens, if the nuclear decay emits several gamma rays in coincidence. If more than one gamma ray is observed by the detector at the same time, it will be treated as one gamma ray with the combined energies. This effect is also known as summing and necessitates that the sample is placed further away, if feasible. With greater distance the efficiency decreases, because the possible incident angles decreases too. \\
Another effect, which is very similar and also dictates to get further away from the detector and keep the rates at a lower level, is called "pile up". Pile up events involve the simultaneous observation of gamma rays from different nuclear decays that just happen to occur at the same time.\par
Measuring short-lived activities can be hard. To get good results a balance between distance, count rate and dead time must be found to capture the activity in time, before it decays away completely.

\chapter{Irradiation}\label{chap:irr} 
Timing can be crucial in these kinds of experiments. Once the samples are irradiated, they have to be transported to the counting lab as quick as possible. Even more importantly, time has to be tracked, to correct for the decays that occur during production and transport. The computer clock in the counting lab was matched with a clock that is used to time the irradiation. All the spectra were recorded with a time stamp from that computer.\par
First of all, the beam was tuned by the operators at the 88-Inch Cyclotron. They make these experiments possible. The ion source, used to produce beam with a certain energy and particle type, must be aligned properly with the cyclotron settings. In use was the AECR, one of three ion sources. The extraction and beam optics, transported the protons over into the cave (see Section \ref{sec:cyc} for details). A glass window with phosphorous coating was aiding the operator in this process. Short bursts of beam lit up the surface that it hit, captured by camera and displayed in the control room. The camera was placed off-axis to the beam, to minimize radiation damage. (See Figure \ref{fig:phosphor} for the setup.)\par
\begin{figure}[htb]
	\centering
	\includegraphics[width=200px, keepaspectratio]{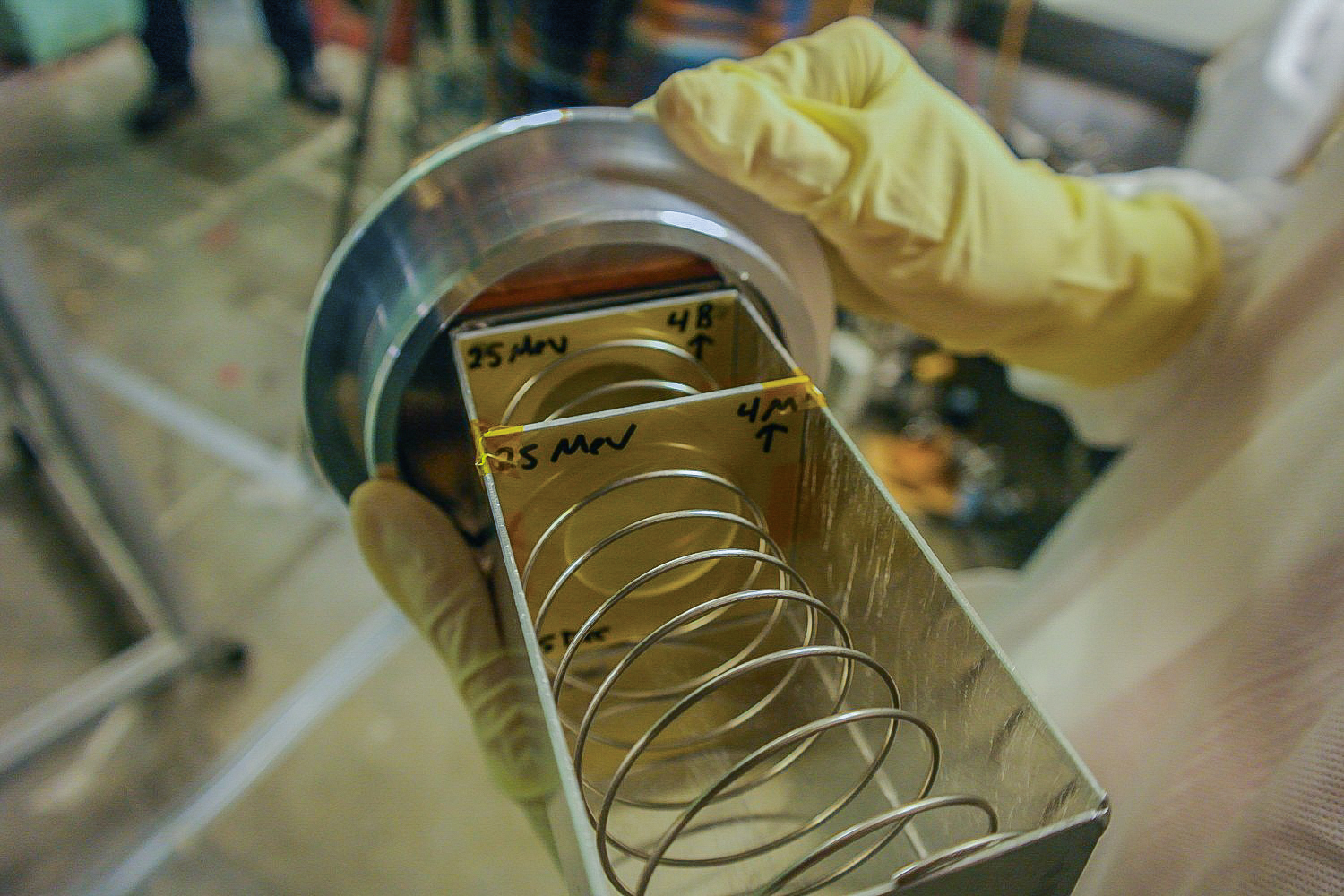} 
	\includegraphics[width=200px, keepaspectratio]{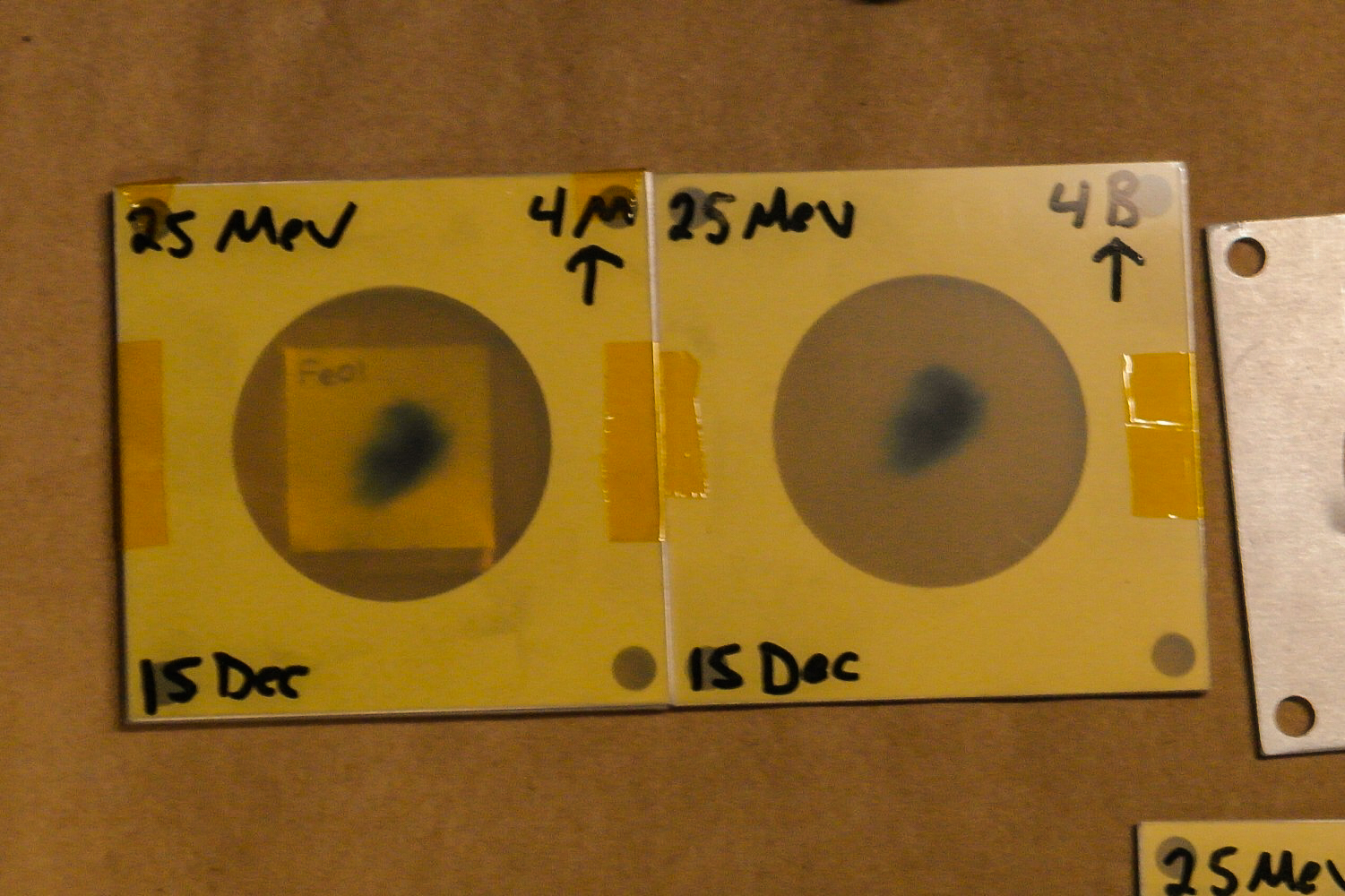}\\
	\caption{Gafchromic film was used to reveal the beam shape. In this case 4M shows the front of the stack and 4B the rear.}
	\label{fig:exp-gaf}
\end{figure}

The next step was to verify that there is indeed good focus of the beam at the stack position in the back of the holder. Therefore, the target holder was prepared with two pieces of Gafchromic film, one at the back and one at the approximated front position of the stack, which changes depending on the thickness of the target stack. These again got hit by the beam for a very short time to expose the film. The 'beam burns' in Figure \ref{fig:exp-gaf} were the last ones taken and show the beam that was used on the low energy stack. 
\par
It took only $30\,\rm{pA}$ of beam for one second, to develop them. These shots built confidence, that the beam under-filled the target and did not change in area or shape in between. The focus therefore was parallel in the stack.\par
It took multiple attempts before a satisfying result was reached, including four beam burns. The decision was made to park the beam over night, so all the time critical work would not have to be performed late at night, while fatigued. Two irradiations were performed at proton energies of 25 and $55\,\rm MeV$ respectively. The irradiation of the $25\,\rm{MeV}$ stack finally got carried out in the afternoon of the 15th of December 2016. During the $20\,\rm{minute}$ irradiation, the current value of the charge integrator was written down every $30\,\rm{seconds}$. This will serve as confirmation later, that the beam current was stable over the course of the experiment. The beam current was set to $100\,\rm{nA}$.\par
Everything else had to be made ready for the subsequent counting process. There were a few short-lived activities, which had to be counted quickly. Namely, the decay of $^{63}\rm{Zn}$ in the copper monitor foils and the direct decay of $^{51}\rm{Mn}$ from the iron foils. They have half-lives of $38.5\,\rm{min}$ and $46.2\,\rm{min}$ respectively. Other activities can be observed in the days and weeks following the experiment.\par
The high energy $55\,\rm{MeV}$ stack followed in the early morning hours of the 16th. Just in time to be done within the requested beam-time window. The process was essentially the same, apart from the fact that the targets were only irradiated for $10\,\rm minutes$ and $31\,\rm seconds$ with $120\,\rm{nA}$ of beam. The beam line, holding the targets, was electrically isolated, allowing for a charge integrator to be used to determine the total number of protons that made incident on the target stacks.  
\begin{table}[hb]
	\centering
	\caption{Summary of target stack and beam parameters}
	\begin{tabular}{ccccc}
		\hline
		Stack & Energy & \# Target packages & Duration & Current \bigstrut[t] \\
		\hline
		Low energy & $25\,\rm{MeV}$ & 7 & $20\,\rm{min}$ & $100\,\rm{nA}$ \bigstrut[t] \\
		High energy & $55\,\rm{MeV}$  & 7 & $10.52\,\rm{min}$ & $120\,\rm{nA}$\\
		\hline
	\end{tabular}
	\label{tab:beam}
\end{table}

\chapter{Results}\label{chap:results} 
The charge integrator read $113.8\,\rm{\upmu C}$ after the low energy stack, and $74.8\,\rm{\upmu C}$ for the high energy stack. That translates into a proton flux of $5.919\cdot 10^{11}\,\nicefrac{\rm{prot.}}{\rm{sec}}$ and $7.399\cdot 10^{11}\,\nicefrac{\rm{prot.}}{\rm{sec}}$ respectively. That is the only directly measured observable, everything else was derived from $\upgamma$-ray spectra, as discussed in depth below.\par
\begin{figure}[htb]
	\centering
	\includegraphics[width=200px, keepaspectratio]{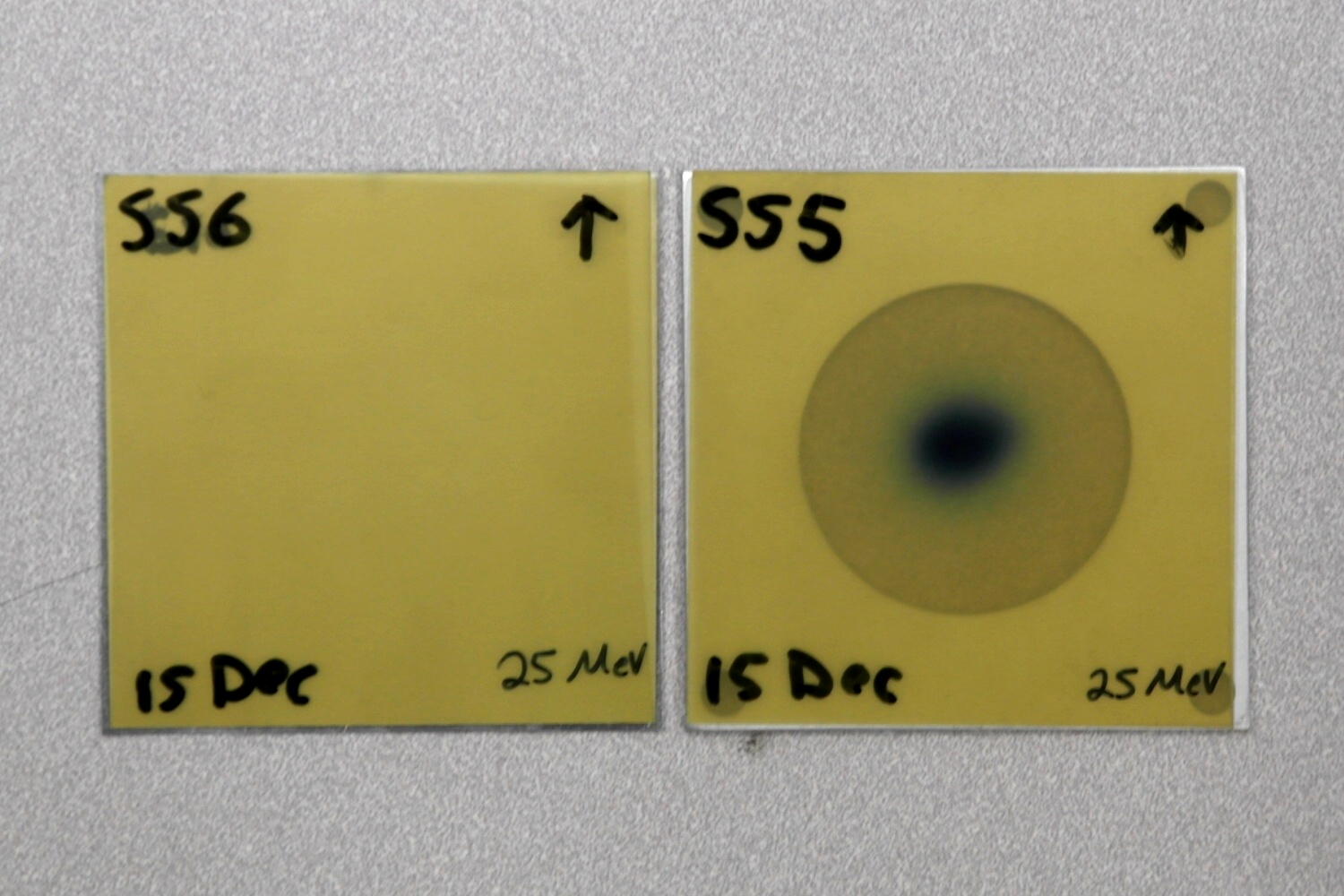} 
	\includegraphics[width=200px, keepaspectratio]{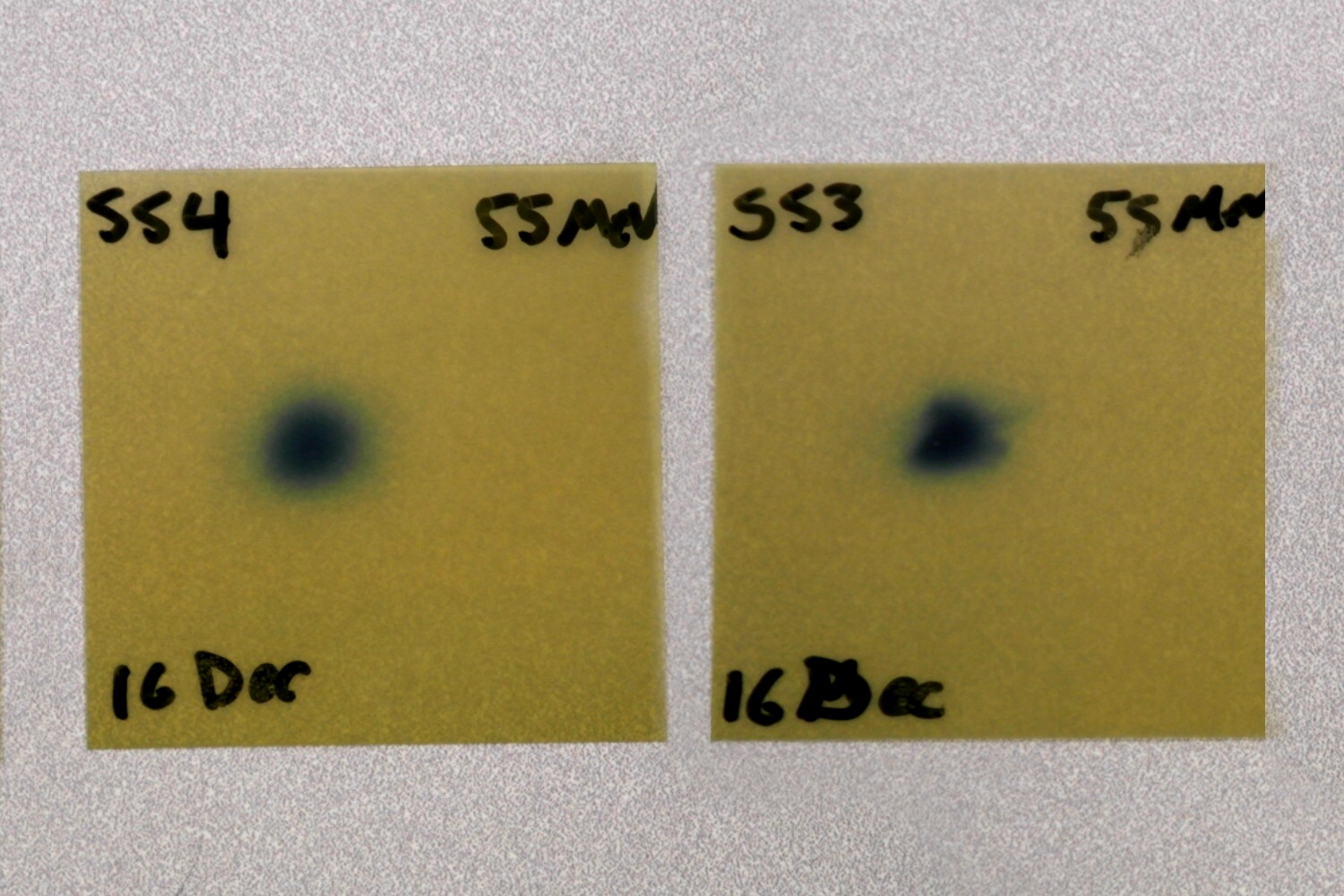}\\
	\caption[Gafchromic film exposed by the stainless steel profile monitors. $25\,\rm MeV$ on the left, and $55\,\rm MeV$ on the right with the front of the stack always on the right. One of the standard frames is shown for reference.]{Gafchromic film exposed by the stainless steel profile monitors. $\mathit{25\,Me\!V}$ on the left, and $\mathit{55\,Me\!V}$ on the right with the front of the stack always on the right. One of the standard frames is shown for reference.}
	\label{fig:ssgaf}
\end{figure}

A couple of hours after the irradiation, Gafchromic film was fixed to the stainless steel plates and sealed lightproof. The film got exposed by the activity of the plates. Figure \ref{fig:ssgaf} shows them for both stacks with the high energy side or front of each stack on the right.
\par
The beam got stopped in the low energy stack, before it could activate the back end profile monitor significantly. This was the first sign of a discrepancy in the estimated beam energy. 

\section{Toolbox}
\begin{figure}[t!]
	\centering
	\includegraphics[width=270px, keepaspectratio]{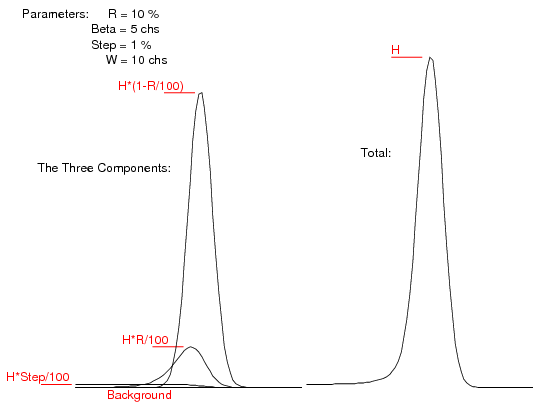}\\
	\caption{The components of the peaks fitted by gf3 \cite{radware-gf3}}
	\label{fig:gf3-peak}
\end{figure}
\begin{figure}[b!]
\centering
	\includegraphics[width=420px, keepaspectratio]{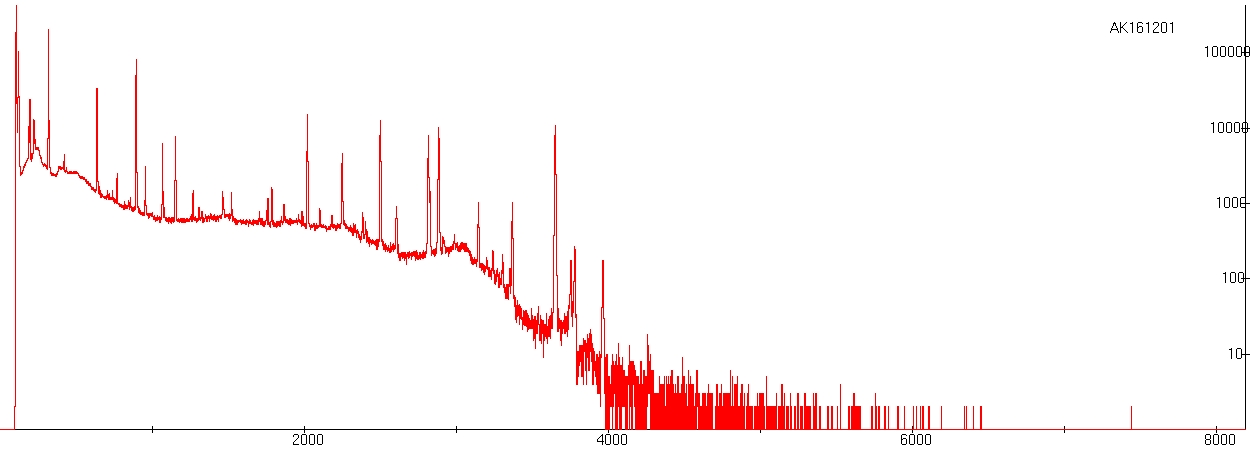}\\
	\caption[Example for a spectrum loaded in gf3, showing the $^{152}\rm Eu$ calibration source $10\,\rm cm$ away from the surface detector. Counts are plotted over channel number.]{Example for a spectrum loaded in gf3, showing the $\mathit{^{152}\!Eu}$ calibration source $\mathit{10\:\!cm}$ away from the surface detector. Counts are plotted over channel number.}
	\label{fig:gf3spec}
\end{figure}

In the acquisition process the gamma ray spectra were handled using Maestro \cite{maestro}. Spectra can be viewed and saved in different formats. To further investigate, another program was used, called RadWare\cite{radware}. Maestro does not save directly to a format, that can be read with RadWare, this is why an internal tool was used for conversion, called \verb|spec_conv| \cite{spec_conv}.\\
RadWare or more precise a part of it called \verb|gf3|, enables the user to navigate the spectra and least-square fit $\upgamma$-ray peaks over background. It is operated via command line, but shows the spectra in a separate window (see Figure \ref{fig:gf3spec}). This tool was invaluable for the analysis. The fit administered, is made up of three components: a Gaussian (main component), a skewed Gaussian and a smoothed step function to increase the background on the low-energy side of the peak. Figure \ref{fig:gf3-peak} illustrates how the three components act together. Multiple peaks can be selected for a given fit, which makes sense, if they are close together, providing a better grasp of the background.
\clearpage
\newpage

\section{Calibration}
\begin{figure}[ht!]
	\centering
	\includegraphics[width=420px, keepaspectratio]{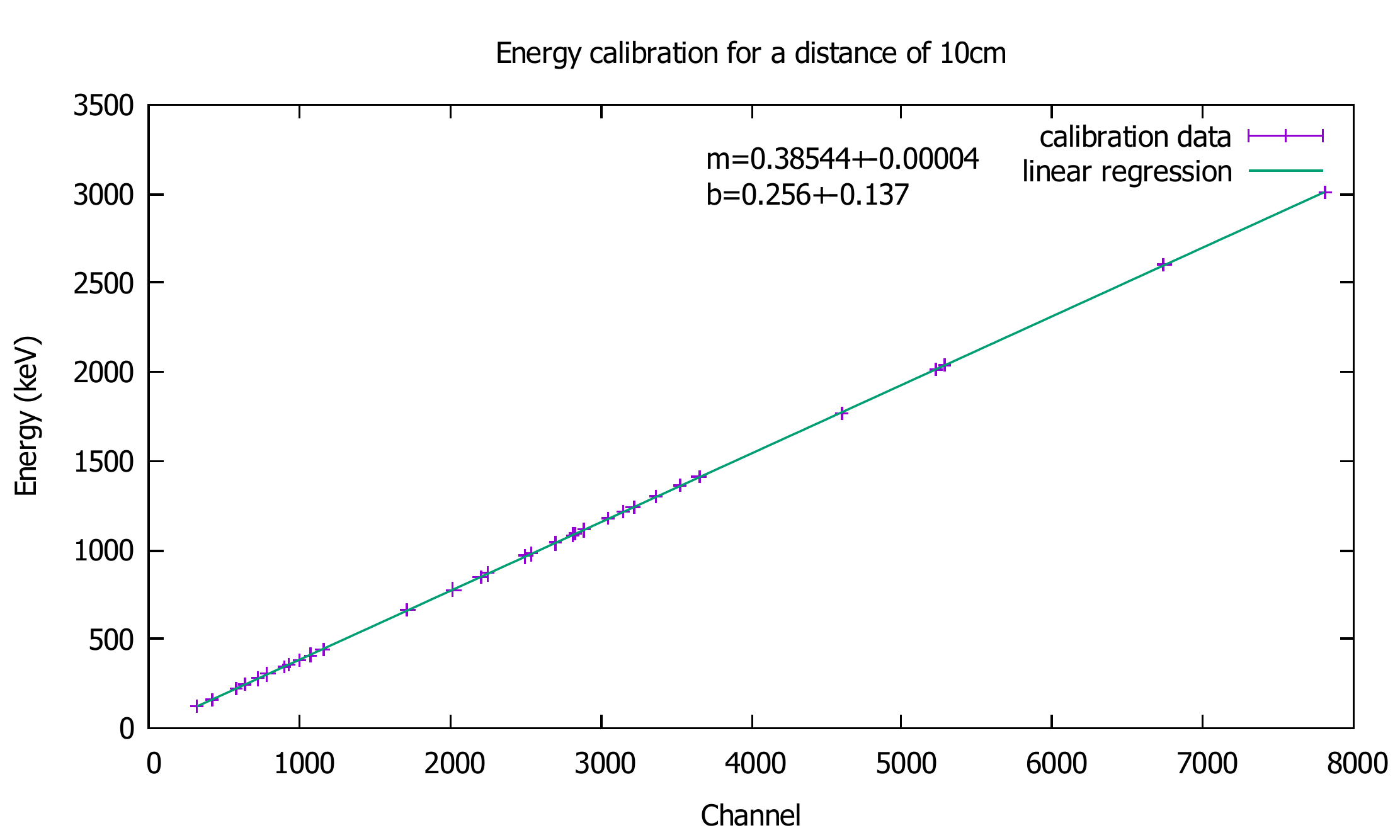}\\
	\caption[Energy curve for $10\,\rm cm$]{Energy curve for $\mathit{10\:\!cm}$}
	\label{fig:ecal}
\end{figure}

Before any kind of analysis could be performed, the germanium detector had to be calibrated. This was accomplished by using calibration sources, which the prominent $\upgamma$-ray energies are very well-known of, as well as the activity that is to be expected. This activity was measured when the calibration source was produced and can be extrapolated to any point in the future via the decay law (Eq. \ref{eq:decay2}).
Calibration spectra had to be taken at every distance, that was going to be used in the counting process. Four different calibration sources were used: $^{133}\rm{Ba}$, $^{56}\rm{Co}$, $^{137}\rm{Cs}$ and $^{152}\rm{Eu}$. Their details are listed in Table \ref{tab:cali}.
\begin{table}[hb]
	\centering
	\caption{Calibration sources used}
	\begin{tabular}{cccc}
		\hline
		Source & Half-life (years) & Activity (Bq) & Reference Date \bigstrut[t] \\
		\hline
		$^{133}\rm{Ba}$ & 10.55 & 39890 & 01/01/09 \bigstrut[t] \\
		$^{56}\rm{Co}$ & 0.21 & 45140 & 12/15/15\\
		$^{137}\rm{Cs}$ & 30.08 & 38550 & 01/01/09\\
		$^{152}\rm{Eu}$ & 13.53 & 370000 & 11/01/84\\
		\hline
	\end{tabular}
	\label{tab:cali}
\end{table}
First the channels $ch$ (bins) from the spectra had to be matched to the photon energies $E_\upgamma$. Every prominent peak (intensity greater than $\approx\!0.5\,\%$), expected from the calibration sources, was used to relate energy to channels. Those were about 32 data points, varying slightly for each position. With a linear regression (example in Figure \ref{fig:ecal}), this relation was put into a functional form (Equation \ref{eq:ecal}). Uncertainties were included, both in energy and channel, but are too small to be visible in the plot. 
\begin{equation}\label{eq:ecal}
	E_\upgamma(ch)=m\cdot ch+b
\end{equation}\par
The primary purpose of the calibration was to allow the peaks of interest to be identified. Once that was done, the listed literature $\upgamma$-ray energies \cite{nndc-nudat} were used for more precision. Therefore, it was very important to correctly identify the peaks.\par
The calibration for efficiency on the other hand, directly effects the measured cross sections. The peak areas were used as measure of how many $\upgamma$-rays were detected, that are associated with the same energy. This was related to the known activity of the source. The different sources must be normalized to account for the difference in activity and counting time. Once the different gamma ray intensities were taken into account, it was possible to calculate an efficiency depended on the energy and distance from the detector. For each distance, a different set of values was obtained. These sets did not show a simple linear dependency. The detector is much better at lower energies and the efficiency $E{\!f\!f}$ declines towards higher energies following a power law: \\
\begin{equation}
	E{\!f\!f}(E_\upgamma) = \exp(p_0 \ln(E_\upgamma) + p_1)
\end{equation}
\begin{figure}[h!]
	\centering
	\includegraphics[width=420px, keepaspectratio]{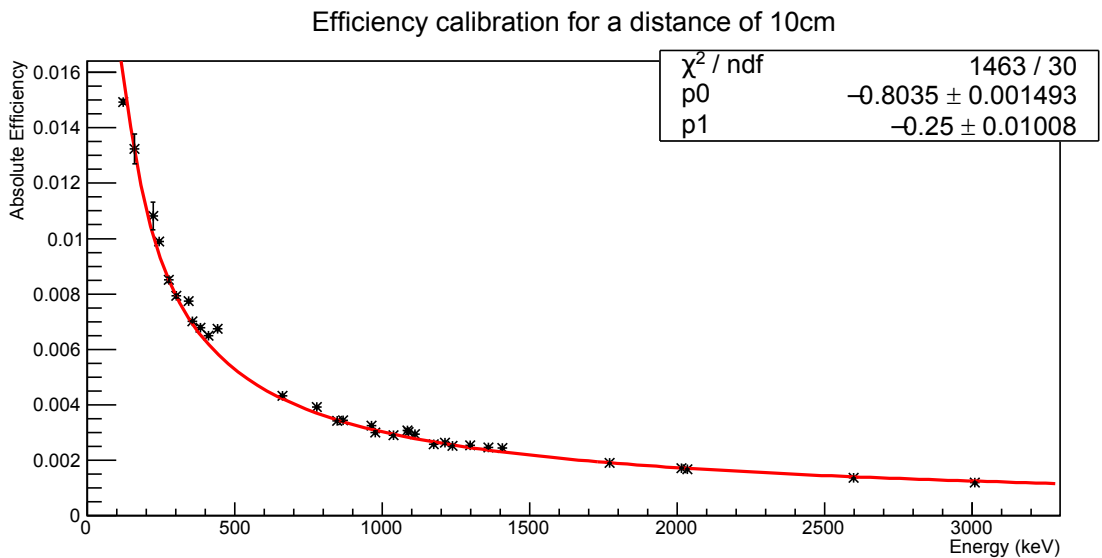}\\
	\caption[Efficiency curve for $10\,\rm cm$]{Efficiency curve for $\mathit{10\:\!cm}$}
	\label{fig:effcal}
\end{figure}\par
For the $10\,\rm{cm}$ position, $\approx\!1.5\,\%$ at low energies and $\approx\!0.2\,\%$ at high energies was detected. With this calibration it was possible to detect and quantify gamma rays with an energy range of $100 - 3000\,\rm{keV}$. For lower energies the efficiency curve would go down again, but since this region was of no concern, the simple power law dependency gives an accurate description. 
\clearpage
\newpage

\section{Analysis}\label{chap:analysis}
A total of 399 spectra were taken between the foils of the two stacks. They were analyzed and the peak areas, as well as their uncertainties, corresponding to the decay of radionuclides of interest were determined, using the \verb|gf3| program. Weak activities were counted first, to get good statistics on them in a reasonable time span. Counts were grouped by foil type, meaning in sets of seven.\par
Every set was stored as separate file to be easily handled in the following analysis. There are a couple of other accompanying parameters that were also put into the same file. As an example, this is what the file of the second copper set in the low energy stack looks like, concerning the production of $^{63}\rm{Zn}$:\par
\begin{table}[h!]
	\caption[Example file for $^{63}\rm Zn$ production on the left, description on the right.]{Example file for $\mathit{^{63}\!Zn}$ production on the left, description on the right.}
	\rule{\textwidth}{1pt}
		\begin{minipage}{9.2cm}
		\vspace{0.1cm}
		\begin{verbatim}
			Zn63 2308.2
			16590	15909	15275	14562	13717	13050	12264
			743	527	538	580	586	513	675
			15	10	10	15	15	5	1
			669.62	0.082	0		
			637	2496	10200	14592	16855	35341	0
			60	76	112	130	140	206	0
			962.06		0.065	0.004
			461	1630	5871	8407	9675	20468	0
			55	63	87	100	107	161	0
		\end{verbatim}
		\vspace{-0.3cm}
	\end{minipage}
	\begin{minipage}{6cm}
		\vspace{0.1cm}
		Nucleus, Half-life\\
		Time since EoB (End of Beam)\\
		Counting time\\
		Counting position\\
		$\upgamma$-line energy, intensity, uncert.\\
		Peak area\\
		Peak area uncertainty\\
		$\upgamma$-line energy, intensity, uncert.\\
		Peak area\\
		Peak area uncertainty\\
		\vspace{-0.3cm}
	\end{minipage}
	\vspace{0.2cm}
	\rule{\textwidth}{1pt}
	\label{tab:file}
\end{table}
\vspace{-0.6cm}
On the right of the example above is a description for each line. Most of the decay is producing $511\,\rm{keV}$ $\upgamma$-rays, but many different decays feed into this line. That is why only characteristic peaks were used. The two most prominent peaks that meet these criteria are $669.62\,\rm{keV}$ and $962.06\,\rm{keV}$ with $8.2\,\%$ and $6.5\,\%$ intensity (branching ratio) respectively. Those were chosen for the analysis of this decay.\par
To make the whole analysis easily adjustable and repeatable, ROOT code was used to perform calculations. Find the code, performing the calculations, in Appendix \ref{app:code}. ROOT is based on the programming language C++ and is widely used in particle physics. Developed and made available by CERN, it comes with a variety of data analysis capabilities. \cite{root}\par
The tabulated data was read in and peak areas adjusted with the efficiency corresponding to the position that the foil was placed at, relative to the detector. They were also adjusted for the different intensities (branching ratio) of the $\upgamma$-rays, to the end that they are comparable. \par
The resulting peak areas were translated into production rates. This required the use of the decay law introduced in Chapter \ref{chap:intro}. The number of decays that occurred during counting, rather than the total umber of nuclei formed, was observed. Equation \ref{eq:decay-n} is used to describe this difference in the number of nuclei, in regard to the rate.
\begin{equation}
	\Delta N = N_2 - N_3
\end{equation}
Where the cooling time for $N_2$, is the time that has passed since the EoB $t_{\rm sEoB}$, and for $N_3$, is that time plus the counting duration $t_{\rm count}$. Combining this and solving for the production rate $R$ results in:
\begin{equation}\label{eq:rate}
	R=\frac{\Delta N \lambda}{1-e^{-\lambda t_{\rm i}}}\left( e^{-\lambda t_{\rm sEoB}} - e^{-\lambda t_{\rm sEoB}+t_{\rm count}} \right)
\end{equation}
The uncertainty in the production rate is directly propagated from the uncertainties contributing to $\Delta N$. Those come from the statistics and the recorded uncertainties in the fitted peak areas. The branching ratio contributes as well, but is small in comparison. Uncertainties in the exponents of Equation \ref{eq:rate}, coming from the decay constant and times, are negligible.\par
\begin{figure}[h!]
	\centering
	\includegraphics[width=400px, keepaspectratio]{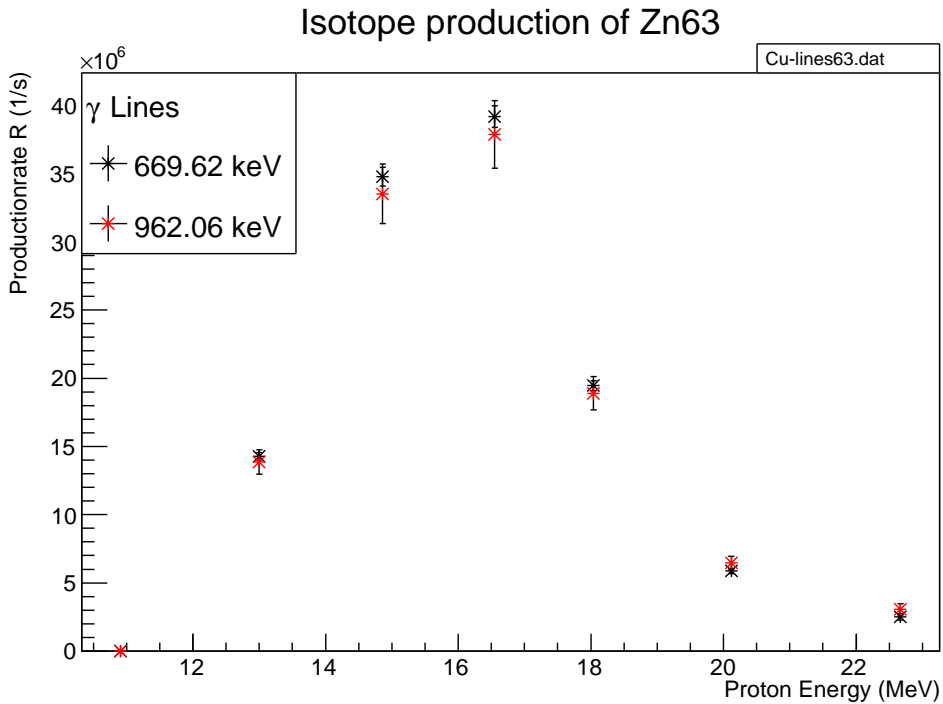}\\
	\caption[Production rate of $^{63}\rm Zn$ (one of two sets), for the low-energy stack]{Production rate of $\mathit{^{63}\!Zn}$ (one of two sets), for the low-energy stack}
	\label{fig:zn63}
\end{figure}
\begin{figure}[h!]
	\centering
	\includegraphics[width=400px, keepaspectratio]{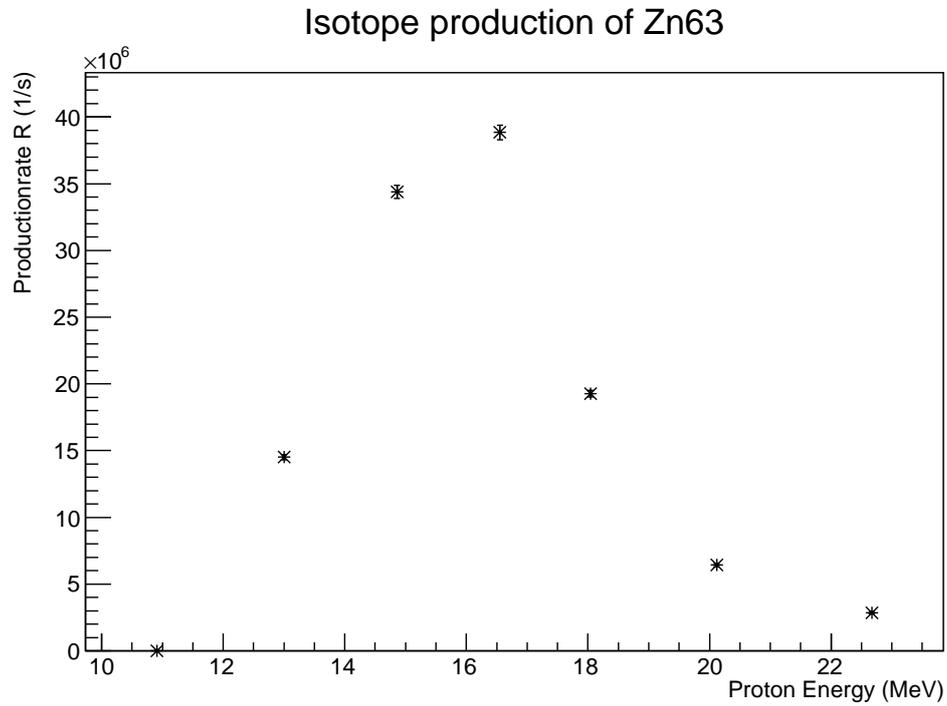}\\
	\caption[Production rate of $^{63}\rm Zn$ weighted average from all sets, for the low-energy stack]{Production rate of $\mathit{^{63}\!Zn}$ weighted average from all sets, for the low-energy stack}
	\label{fig:zn63-tot}
\end{figure}

For most production rates, there were at least 2 sets used with two lines each, resulting in 4 data points per foil. See Figure \ref{fig:zn63} for an example of one set with 2 lines for the production of $^{63}\rm{Zn}$, which is one of the monitor reactions used. Putting all data points together, reduces the uncertainty from the counting and peak fitting process. The final plot of the combined rates for $^{63}\rm{Zn}$ shows that in Figure \ref{fig:zn63-tot}.\par
\begin{figure}[b!]
	\centering
	\includegraphics[width=420px, keepaspectratio]{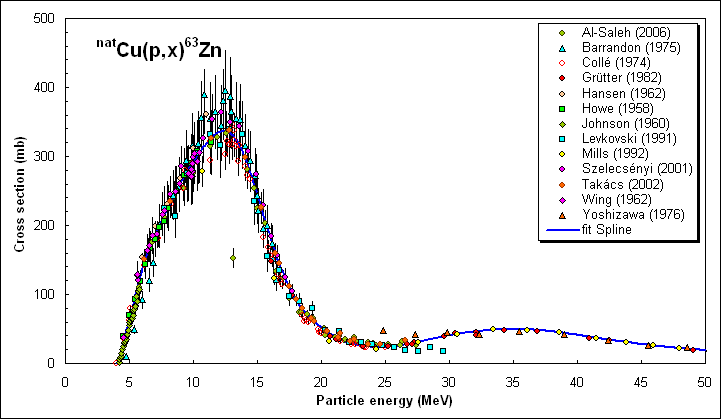}\\
	\caption[IAEA recommended cross section for the $\rm ^{nat}Cu(p,x)^{63}Zn$ reaction \cite{iaea}]{IAEA recommended cross section for the $\mathit{^{nat}\!Cu(p,x)^{63}\!Zn}$ reaction \cite{iaea}}
	\label{fig:zn63-iaea}
\end{figure}
\begin{figure}[bt]
	\centering
	\includegraphics[width=420px, keepaspectratio]{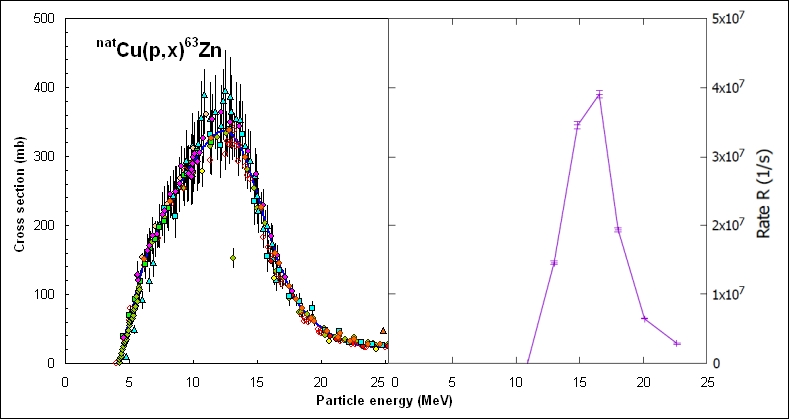}\\
	\caption{IAEA recommended cross section \cite{iaea} compared to the experiment, Figure \ref{fig:zn63-tot}}
	\label{fig:splitzn63}
\end{figure}

\newpage
On the x-axis on both of these plots, the beam energy is shown, that was expected those foils to experience. A comparison between the monitor reaction production rates with the IAEA recommended cross sections (Figure \ref{fig:zn63-iaea} \cite{iaea}) shows a systematic shift in energy. (see Figure \ref{fig:splitzn63}). This issue was a recurring feature of all charged particle stacked target experiments performed at the cyclotron in 2016 and 2017. Without knowing the energies, the production rates of the monitor channels can not be turned into proton fluxes. There was a reading on the proton flux from the charge integrator as well, but any escaping charge carrier would influence this value. It is always favorable to get the flux from the monitors, because those are usually the more trustworthy.\par
Last but not least, the energy is needed to report the energy differential cross section of the main reaction. Without correct energy values the result is less valuable.\par 

\subsection{Empirical Model}
In order to represent the problem, an empirical model was constructed from the monitor channels. One of the assumptions made was, that the flux $\Phi$ was constant throughout the stack. This will hold true, as long as the product of the target areal density and total reaction cross section is much less than unity, which is the case for these experiments. The other assumption was, that the foils were identical (number of target nuclei $N_{\rm T}$ was constant), which is justified, since they have the same thickness. This assumption was only made here. Anywhere else, the actual areal density was used. The rates for one channel were multiplied by the same factor to match the magnitude of the cross section. This comparison can be made because of the two assumptions and the fact, how rate $R$ and cross section $\sigma$ relate to each other, introduced in Equation \ref{eq:cs}. Solving for $R$ gives:
\begin{equation}
	R= N_T\cdot \Phi\cdot \sigma
\end{equation}
\begin{figure}[t!]
	\centering
	\includegraphics[width=400px, keepaspectratio]{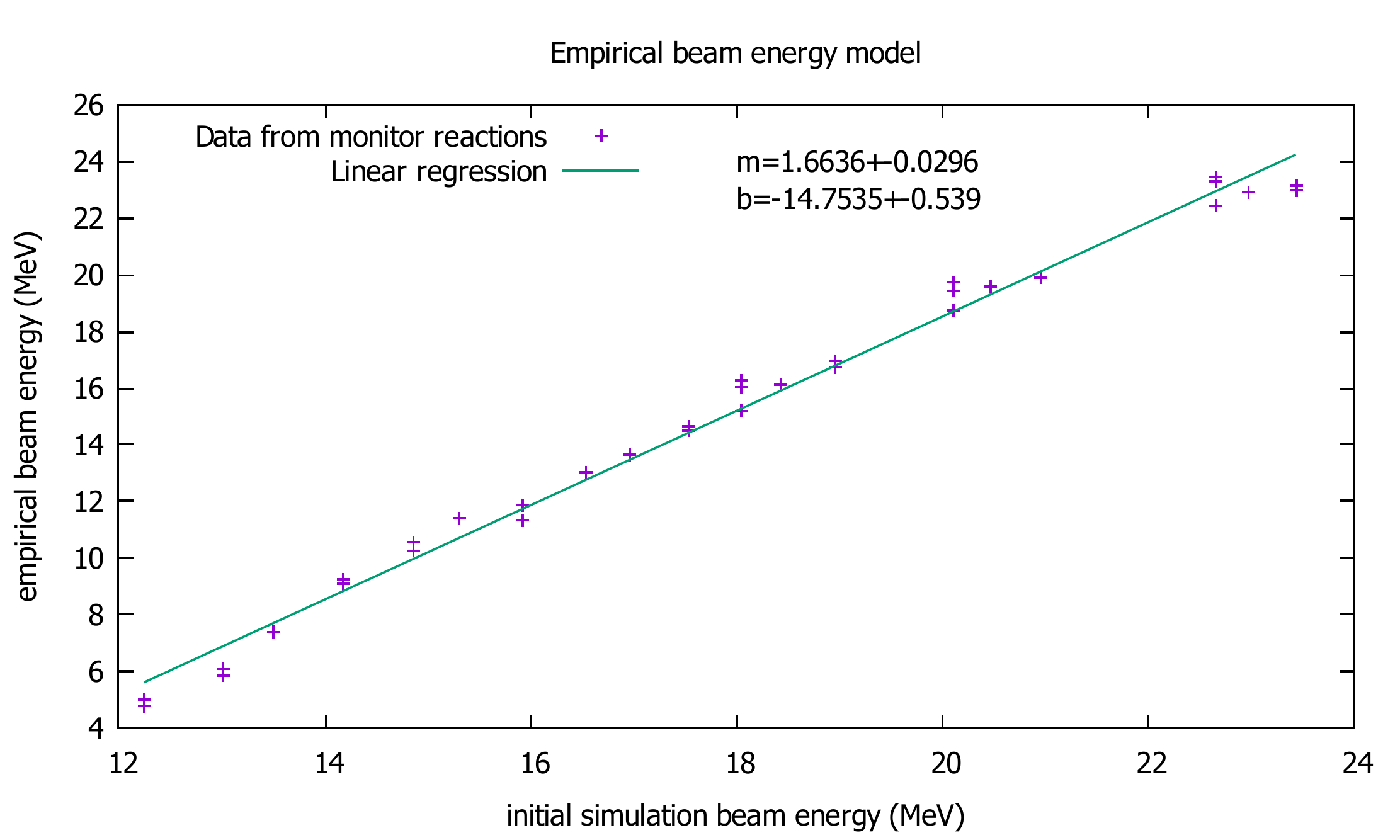}\\
	\caption{Empirical model to relate the calculated energies to the observed production rates in the monitor foils}
	\label{fig:emp}
	\vspace{-0.5cm}
\end{figure}
\begin{figure}[b!]
	\centering
	\vspace{-0.5cm}
	\includegraphics[width=400px, keepaspectratio]{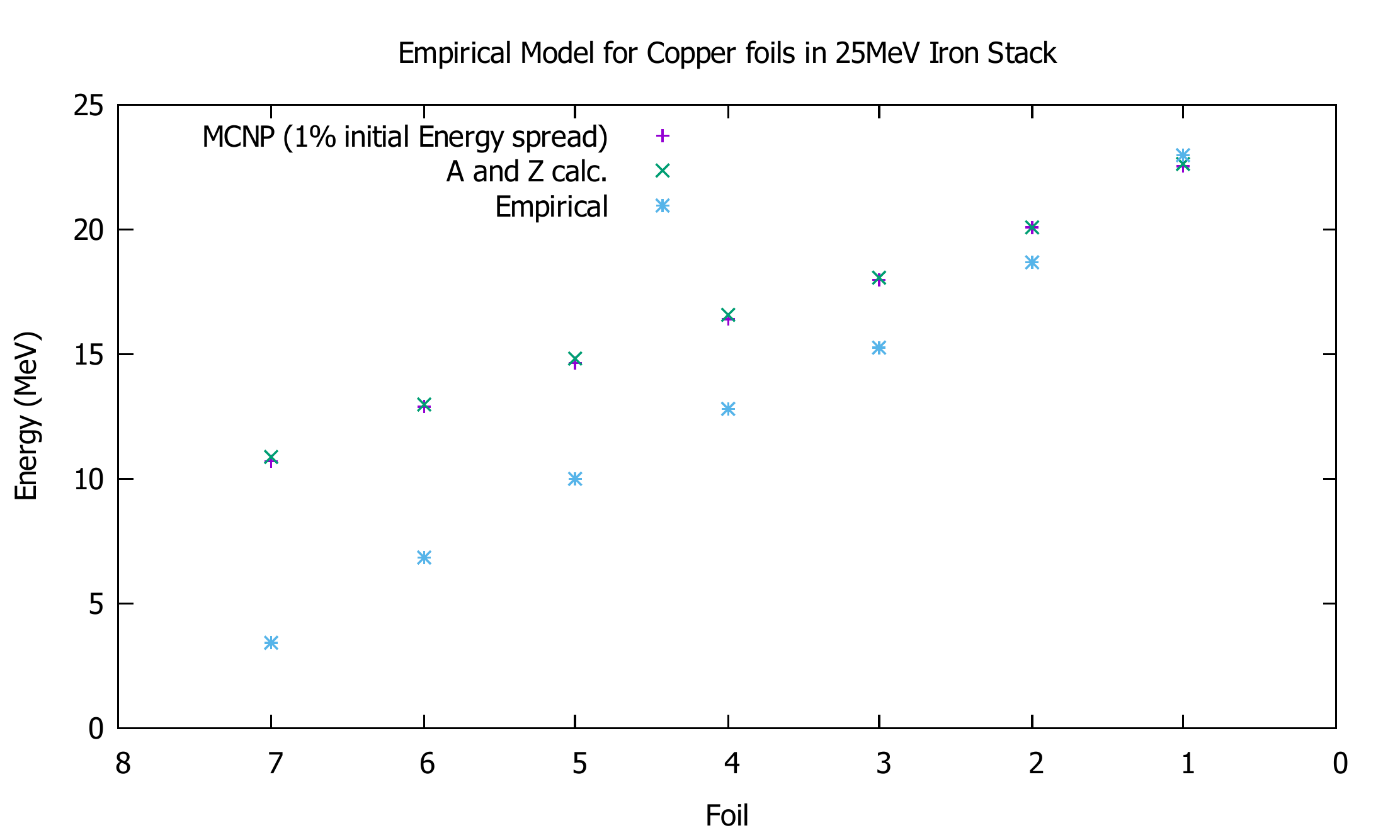}\\
	\caption{Empirically found energies do not agree with the other models}
	\label{fig:sim-emp}
\end{figure}\ \ \ \
The energies for each measurement were then adjusted to best fit with the shape of the cross section curve. When this was done for all the monitor channels, the old result was compared with the new estimate and a function determined to describe the behavior. This empirical model is shown in Figure \ref{fig:emp}. The relation is linear, and a polynomial does not result in a better fit. The linearity of the effect suggests that the problem is not due to a single foil, but rather to the entire stack. It was consistently underestimated how much energy is being lost throughout the stack, as Figure \ref{fig:sim-emp} shows.\par
After assessing many potential causes, it was discovered, that the $25.4\,\upmu\rm m$ thick kapton tape (visible in Figure \ref{fig:det}) used to hold all the iron and monitor foils, also had a $38.1\,\upmu\rm m$ acrylic adhesive, that caused almost as much energy loss in every foil as the kapton tape itself \cite{kapton}. This explains a lot of what was going on and might be the most valuable realization of this work.

\subsection{MCNP Model}
To solve this issue, a more sophisticated simulation is needed. The solution lies in a matrix of simulations, where the composition of the adhesive, initial beam energy and spread, are varied slightly until the best fit to the monitor data is obtained. Using this approach, it was determined that, the initial beam energy is likely to be a minor factor. More on that can be found in the Appendix \ref{app:spread}.\par
\begin{figure}[b!]
	\centering
	\includegraphics[width=400px, keepaspectratio]{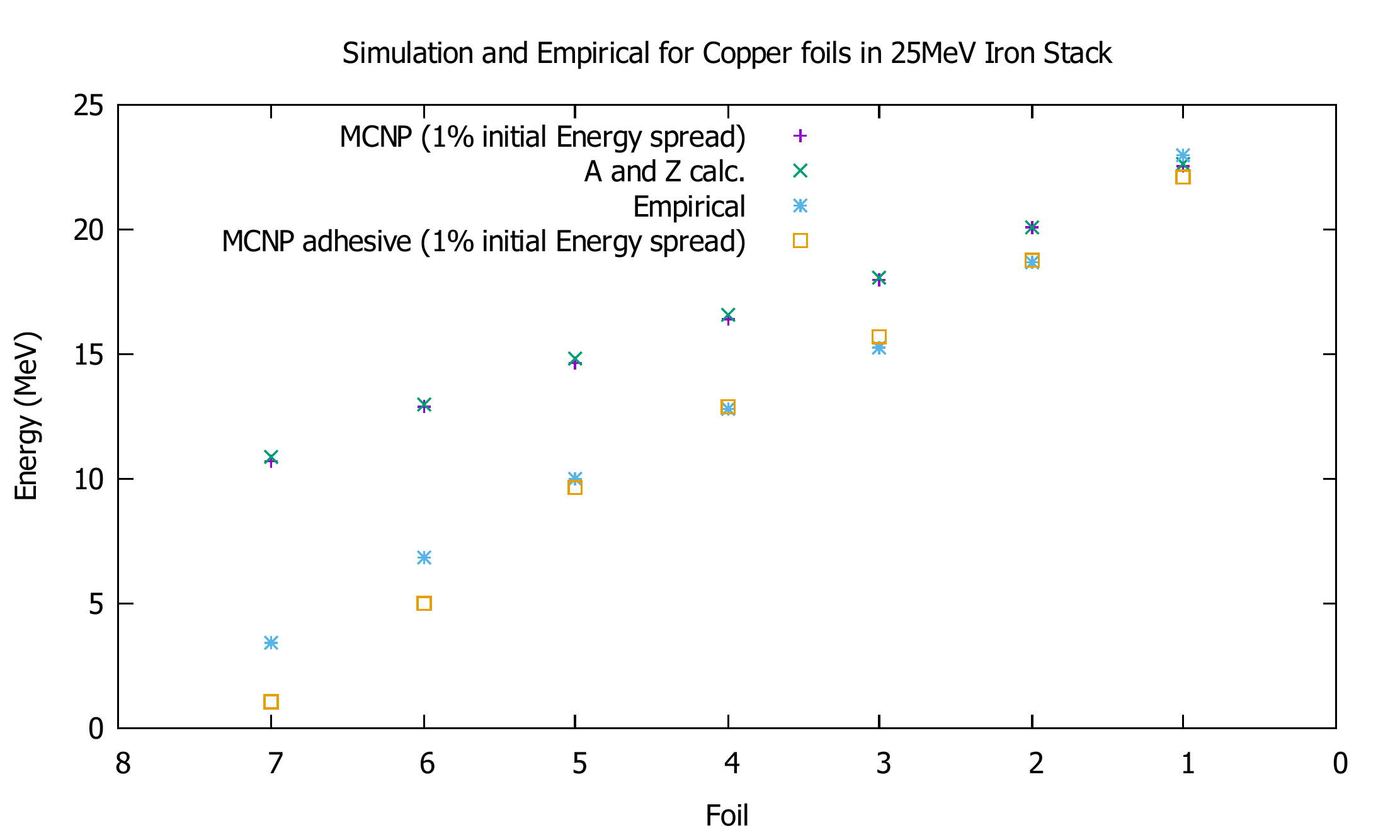}\\
	\caption{Encouraging agreement between empirical model and MCNP simulation}
	\label{fig:sim-emp-sim}
\end{figure}

\begin{figure}[bt]
	\centering
	\includegraphics[width=350px, keepaspectratio]{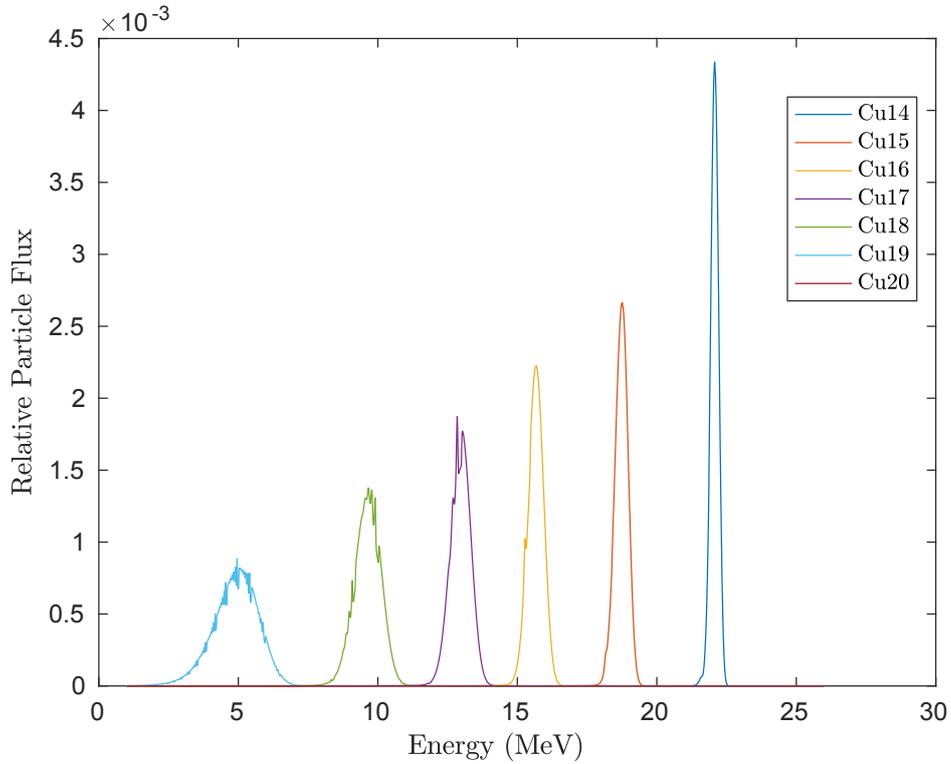}\\
	\caption{MCNP proton flux in the copper foils of the low energy stack. The flux actually only decreases slightly, but the energy distribution widens. Find the log plot in the Appendix \ref{app:flux-log}, Figure \ref{fig:flux-log}}
	\label{fig:flux}
\end{figure}

 
The code used for the simulation is MCNP (Monte Carlo N-Particle Transport Code) \cite{mcnp}, developed and maintained by the Los Alamos National Laboratory. Taking into account the energy loss in the acrylic adhesive, showed promising results right away. In Figure \ref{fig:sim-emp-sim} the comparison was made to the empirical model, and the previous values for the copper foils. The same was done for titanium. The MCNP simulation agrees very well, although some discrepancies are apparent toward the end of the stack, due to the rapid $\nicefrac{{\rm d}E}{{\rm d}x}$ at low energies.
Since MCNP is a Monte Carlo simulation, it gives detailed information about every individual projectile. Once the simulation was computed, the flux of protons as a function of energy $\phi (E)$ (see Figure \ref{fig:flux}) was convolved with the cross section of that monitor reaction $\sigma (E)$ (like Figure \ref{fig:zn63-iaea}). For that to work the IAEA cross section was pade-fit \cite{pade} again, to be able to break it up into smaller bins. It is the same process the IAEA uses to produce their table, except that the bin size is much larger. The other distinct advantage of fitting the cross section again, would have been to extract uncertainties as well. Unfortunately the IAEA cross sections have no stated uncertainties.\\
Including the areal densities of the thin target foils $\rho \Delta r$ as well, and it was possible to calculate the production rate $R'$ that would have been expected, being the integral over all of this:
\begin{equation}\label{eq:Rprime}
	R'=\int \sigma (E)\phi (E) \rho \Delta r \, {\rm d} E
\end{equation}
Since very small bin widths were used, the integral can be approximated with the sum over all energies. In Equation \ref{eq:X} the actual measured rate $R$ is compared to $R'$. The factor $X$ between them, is a scaling factor. The goal is to minimize the spread in $X$. Especially when looking at the $X$ values of the other foils in the same stack, $X$ should be constant.\par
\begin{equation}\label{eq:X}
	X=\frac{R}{\sum_E \sigma (E)\phi (E) \rho \Delta r}\approx \frac{R}{R'}
\end{equation}
\par
\begin{figure}[b!]	
	\centering
	\includegraphics[width=400px, keepaspectratio]{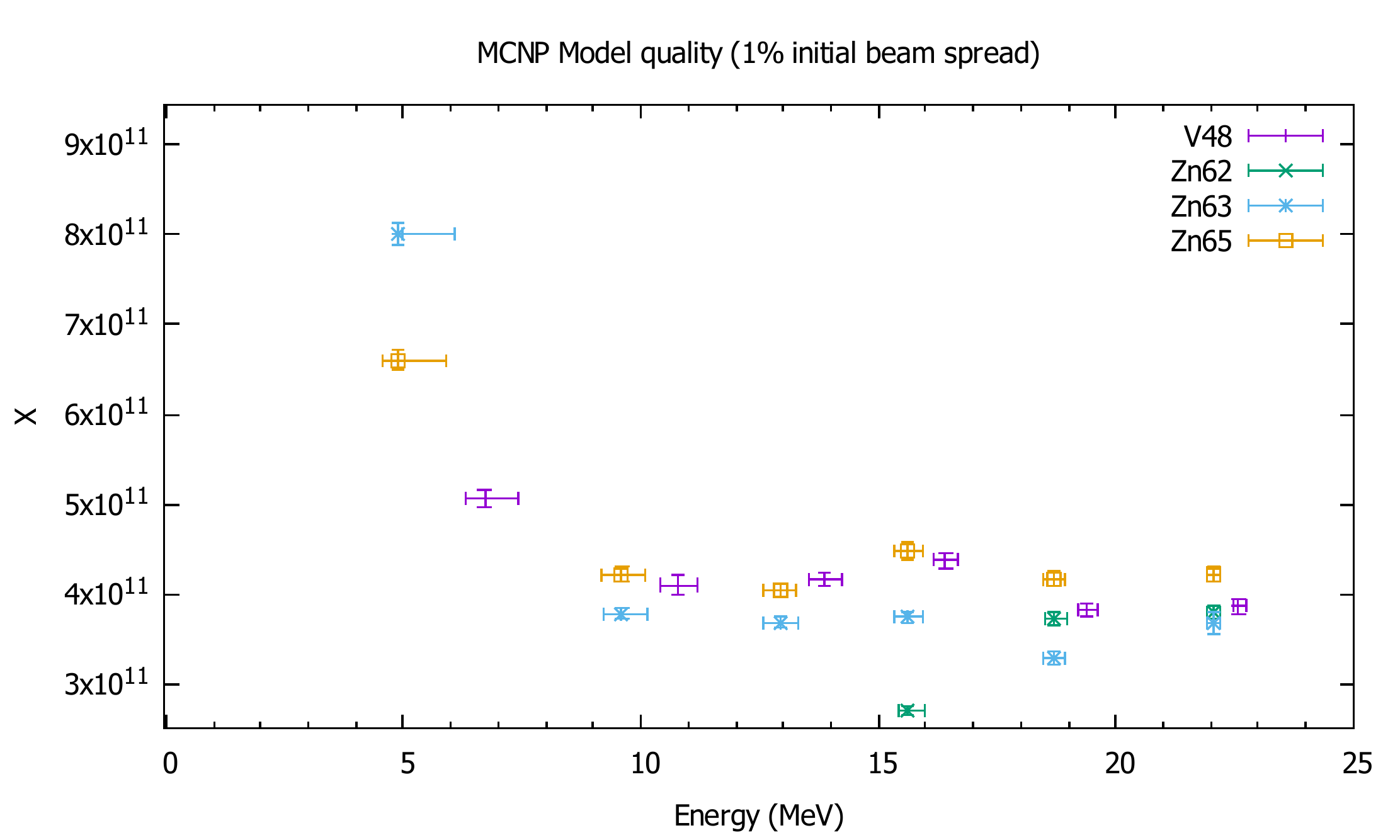}\\
	\caption{X values comparing the different monitor channels. The discrepancy at the end of the stack is clearly visible. For the most part X is constant. Find a cropped view, excluding the outliers, in Appendix \ref{app:xlin} Figure \ref{fig:xlin}}
	\label{fig:x}
\end{figure}

Looking at $X$ allows fine-tuning of the MCNP model. The energy drop towards the end of the stack (which is too steep according to the empirical model in Figure \ref{fig:sim-emp-sim}), is  mirrored in the $X$ value in Figure \ref{fig:x}, which is going up. At the threshold of the reaction, the cross section falls off abruptly. $X$ and therefore the model, becomes very sensitive (even unstable) at that point.\\
Only the uncertainties from the experiment ($R$) were taken into account in Figure \ref{fig:x}. \par
In Figure \ref{fig:x}, and for that matter in general when using the MCNP model, the energy centroid for each foil is defined as the flux weighted average energy of that foil. 
\begin{equation}
	E_{\rm centroid}=\frac{\sum\Phi (E) \cdot E}{\sum \Phi(E) }
\end{equation}
For the energy uncertainty of the monitor foils, the shape of the cross section was taken into account. Flux and cross section were convolved and the uncertainty defined as one sigma around the average, $34.1\,\%$ above and below, resulting in asymmetric error bars (see Appendix \ref{app:flux.c} for detailed proceeding). The final energy centroid of the iron foils, to be reported in the $^{51}\rm{Mn}$ production cross section, was defined in the same way. The uncertainty is different however, since there is no given cross section, the one sigma error was assumed around the flux weighted average.
\clearpage
\newpage

\subsection{Retrieving primary cross section}\label{sec:get}
Two approaches, by which to observe the production of $^{51}{\rm Mn}$, were conceived of. The first and most obvious would have been via the direct decay of the radionuclide. Unfortunately, one of the isotopes best properties, the low direct gamma emission, makes it hard to witness the decay. There is no clean $\upgamma$-line that could have been used. There is however the strong $511\,\rm keV$ line, from the positron annihilation following the $\upbeta$-decay. Because of their origin, these photons are likely to have a detection efficiency different from the one determined using point calibration sources. Since many radionuclides produce $511\,\rm keV$ $\upgamma$-rays, it is difficult to distinguish them from each other. It is only possible if their decay constants differ enough from each other. The way then would be to measure the decay at many points in time and that way record the cumulative decay curve. It would show distinct changes in the slope, whenever the decay of the radionuclide with the shortest half-life dies off and contributes less and less. By fitting these different decay curves to the measured one, the different isotopes could be separated.\par 
The attempt was made in this work, but too few spectra were taken right after the irradiation. This was due to the capabilities of the spectroscopy lab, and an expansion is considered for the future, to enable parallel counting.\par
In the second method however, the observed decay is that of the granddaughter $^{51}\rm Cr$. 
\begin{equation}
	^{51}{\rm Mn} \rightarrow \phantom{}^{51}{\rm Cr} \rightarrow  \phantom{}^{51}{\rm V}
\end{equation}
Only a single prominent and uncontaminated $\upgamma$-line, with an energy of $320\,\rm keV$, is to be expected and was observed. It took some time before enough $^{51}\rm Cr$ was produced to see the $\upgamma$-line. With its $27.7\,\rm day$ half-life however, there was enough time to capture the activity. This method was used to produce the preliminary result presented in Chapter \ref{chap:conc} Figure \ref{fig:preresult}.
\newpage
\leavevmode\thispagestyle{empty}\newpage

\chapter{Conclusion}\label{chap:conc} 
The MCNP simulation and the minimization of spread in $X$ are ongoing. The code will have to run hundreds of times on a computer cluster in order to do that. The most important, and initially unexpected, conclusion, is the effect that the kapton tape adhesive has on the beam energy as the stack is traversed. This is especially important for lower energy stacked target experiments, but would still contribute at higher energies, only it would be more difficult to discover.\par 
This work is similar to a recent study by Graves et al. \cite{graves} on iron, copper and aluminum between 35 and $90\,\rm{MeV}$. Graves et al., also encountered a faster than expected, decrease in beam energy. In their case, it was attributed to a difference in the areal density of the aluminum degraders that were used. The areal density of the individual degraders were varied in an MCNP simulation, until it best reproduced the results that were indicated by the monitors.\par
Graves et al., did not publish a cross section for the production of $^{51}\rm{Mn}$, although they did report results for other exit channels \cite{graves}. Since the energy regions of this work overlap with Graves work, it will be possible to compare the other reaction channels once they are known from this experiment.\par
The method of determining the true beam exposure and energy, will be used in other stacked target experiments, like one that was conducted in March 2017 at LBNL. This was an experiment, utilizing zirconium targets in two stacks and a deuteron beam, to measure the production of medical yttrium isotopes. It will also be used in future efforts, including a planned experiment to measure the production of neutron-deficient cerium nuclei formed in $139\rm La(p,xn)$ reactions.

\section{Preliminary Result}
It is possible to obtain a preliminary $^{51}\rm{Mn}$ production cross section at this time. While the monitor reaction data can not be turned into a correct proton flux yet, the reading from the charge integrator may be used to determine the beam current. The resulting proton flux (low energy stack) is $5.919\cdot 10^{11}\,\nicefrac{\rm{prot.}}{\rm{sec}}$ as stated right at the beginning of Chapter \ref{chap:react}. Likewise, the beam energies without the MCNP model completed, can be substituted by the empirical model values.
\begin{figure}[t!]	
	\centering
	\includegraphics[width=435px, keepaspectratio]{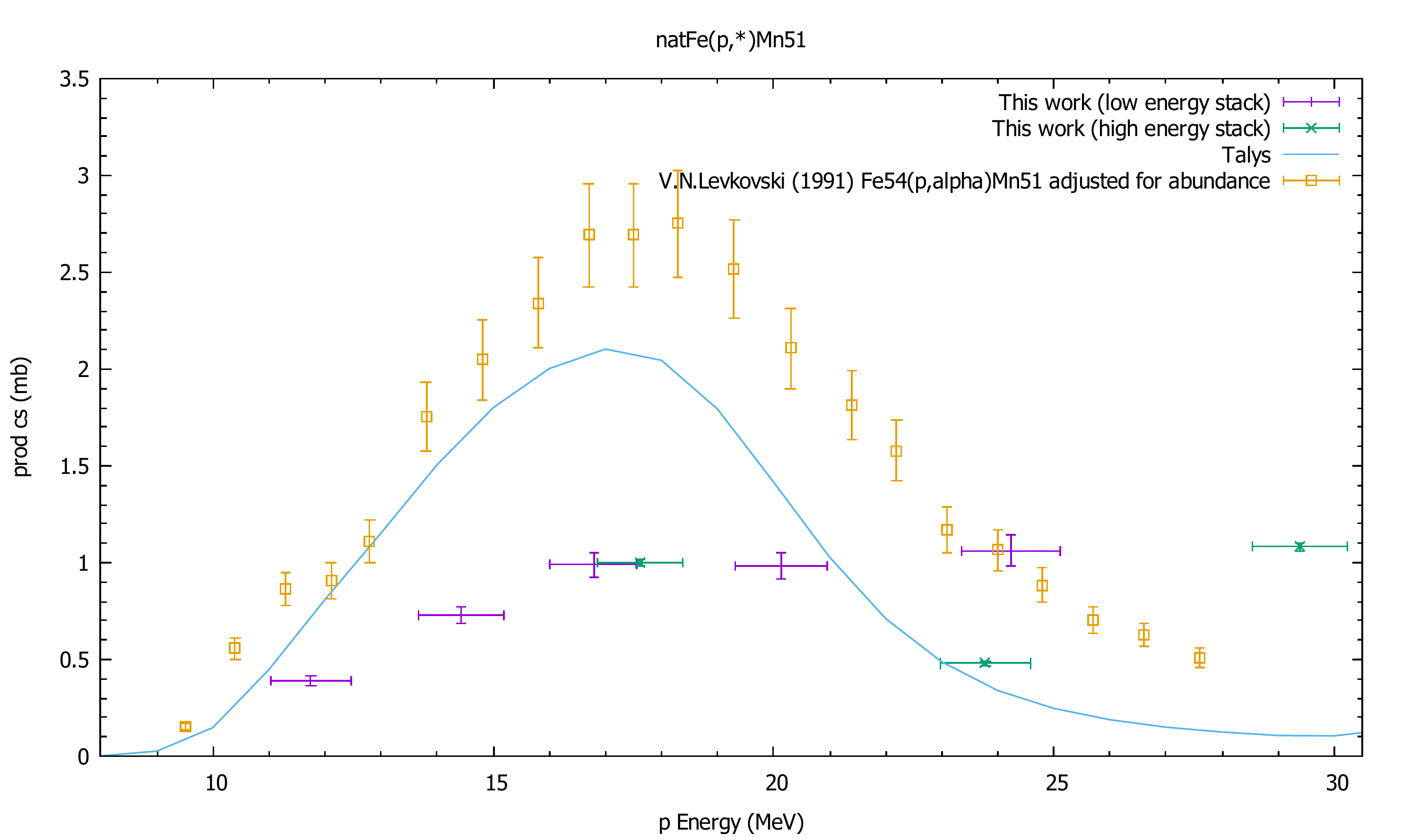}\\
	\caption[Preliminary $^{51}\rm Mn$ production cross section.]{Preliminary $\mathit{^{51}\!Mn}$ production cross section.}
	\label{fig:preresult}
\end{figure}
Figure \ref{fig:preresult} shows the $^{\rm{nat}}\rm{Fe}(\rm p,x)^{51}\rm{Mn}$ cross section, as a function of proton energy, in comparison to the earlier results from Levkovski. The high and low energy stack values overlap visibly. A mismatch between the two can be attributed to an incorrect proton flux and incorrect energies used in this plot. Nonetheless, the cross section falls short of the Talys prediction and the only other available data by Levkovski \cite{lev} as listed on EXFOR (reaction data library) \cite{exfor}. Talys only ever gives a gauge on the possible cross section, and the underlying optical model might even draw from the Levkovski data. This data was later renormalized down and uncertainties included, because of issues with the interpretation of the monitor data in the original report \cite{otuka}. In order to assess whether there is a problem with either the Levkovski data or this work, it is paramount to complete the MCNP model.\par 
In the high energy region there is no cross section determined. This is because, at the threshold of about $30\,\rm{MeV}$, the direct production of $^{51}\rm{Cr}$ opens up and is inseparably entangled with the granddaughter of the main reaction. The direct production is much stronger and drowns out the weaker contribution from the $^{51}\rm{Mn}$ decay chain. It would have been necessary to observe the difference in the decay curves over time, but no early enough measurement of the $^{51}\rm{Cr}$ line was taken. The foils have been counted, but not long enough to pick up the faint signal. Something to consider for future experiments.

\section{Outlook}
As to the future of this program, there are many more production cross sections of medical and other interests, already listed. In May 2015 the United States Nuclear Data Program (USNDP) held a workshop on Nuclear Data Needs and Capabilities of Applications (NDNCA) at LBNL. The goal of the workshop was to compile nuclear data needs across a variety of applications. A whitepaper was produced, summarizing the workshop and pointing out data needs. It is also listing all the specific nuclear reaction cross sections of interest in its Appendix B \cite{whitepaper}.
This comprehensive list is a good resource, when conceiving of future experiments. Considering the length of this list, stacked target experiments are likely to be continued at LBNL.
\newpage
\leavevmode\thispagestyle{empty} 

\newpage
\leavevmode\thispagestyle{empty}\newpage

\renewcommand\thechapter{\Alph{chapter}}
\setcounter{chapter}{1}
\setcounter{section}{0}

\chapter*{Appendix}
\addcontentsline{toc}{chapter}{Appendix}
\renewcommand{\chaptermark}[1]{\markboth{#1}{}}
\chaptermark{Appendix}

\section{Implementation of production rate formula in ROOT macro}
\label{app:code}
\begin{verbatim}
// ************************************************************ //
// Calculations! 
// ************************************************************ //
// Decays detected
Double_t decays[nlines][nfoils];	Double_t decays_err[nlines][nfoils];
// Production rates
Double_t R[nlines][nfoils];			Double_t R_err[nlines][nfoils];
l=log(2)/THalf; // Decay constant
lm=log(2)/2772; // Decay constant for Mn 51 with half-life 46.2 min.
//lm=log(2)/2736; // Decay constant for Mn 51 with new half-life
//of 45.6 min. No significant difference here.
	
for (int i=0; i<nlines; i++){
 Double_t E = E_[i];
 Double_t I = I_[i];
 Double_t Ierr = Ierr_[i];	
 for (int j=0; j<nfoils; j++){
  Double_t Eff, Efferr;		
  Double_t p0 = p0_[codepos_[j]];	Double_t p0err = p0err_[codepos_[j]];
  Double_t p1	= p1_[codepos_[j]];	Double_t p1err = p1err_[codepos_[j]];
  // Eff = efficiency of the detector
  Eff = exp(p0*log(E)+p1);
  Efferr = sqrt(  pow(log(E)*pow(E,p0)*exp(p1)*p0err,2)
    +  pow(pow(E,p0)*exp(p1)*p1err,2)  );
  // Att = mass attenuation coefficient mu/rho in aluminum
  Double_t Att;
  Att = 0.0608857*pow(E,-0.442437);
  printf("Att-expo-factor: %g\n",exp(Att*2.7*0.5));
  Double_t TsEoB = TsEoB_[j];
  TT[file][j] = TsEoB;
  Double_t TLive = TLive_[j];
  // A=Area in the peak. = Number of counts in that peak. 
  Double_t A = Foils[i][j];	Double_t Aerr = Foilserr[i][j];
		
  // Number of decays in the observed time window
  // = \Delta N = N(TsEoB)-N(TsEoB+Tlive)
  decays[i][j] = A/( Eff * I );
  decays_err[i][j] = sqrt(  pow(Aerr/(Eff*I),2)
     +  pow(A*Efferr/(pow(Eff,2)*I),2)
     +  pow(A*Ierr/(Eff*pow(I,2)),2)  );
  // modified with attenuation in the aluminum shield:
  if (alsandwich){
   decays[i][j] = A/( Eff * I ) * exp(Att*2.7*0.5);
  }		
  // Production rates R in the case of decay to the daughter
  R[i][j] = decays[i][j] * l / (1-exp(-l*TIrr)) 
     *   1/(exp(-l*TsEoB)-exp(-l*(TsEoB+TLive)));
  R_err[i][j] = decays_err[i][j] * l / (1-exp(-l*TIrr))
     *   1/(exp(-l*TsEoB)-exp(-l*(TsEoB+TLive)));		
		   
  // Production rate R when the decay radiation originates 
  // from the granddaughter 51Cr
  if (granddaughter){ // l = lc = decay constant for Cr 51
   printf("\n Only for processing 51Cr decay data. \n");
   R[i][j] = decays[i][j] * pow((exp(-l*TsEoB) - exp(-l*(TsEoB+TLive)))
     *   ( 1/l + exp(lm*TIrr)/(lm-l) - (lm*exp(-l*TIrr))/(l*(lm-l)) )
     +   ( 1 - exp(-lm*TIrr) )
     *   ( (exp(-lm*TsEoB)-exp(-lm*(TsEoB+TLive))-exp(-l*TsEoB)
     +exp(-l*(TsEoB+TLive)))/(l-lm) )
    ,-1);
													
   R_err[i][j] = decays_err[i][j] * pow((exp(-l*TsEoB)
     - exp(-l*(TsEoB+TLive)))
     *     ( 1/l + exp(lm*TIrr)/(lm-l) - (lm*exp(-l*TIrr))/(l*(lm-l)) )
     +     ( 1 - exp(-lm*TIrr) )
     *     ( (exp(-lm*TsEoB)-exp(-lm*(TsEoB+TLive))-exp(-l*TsEoB)
     +exp(-l*(TsEoB+TLive)))/(l-lm) )
    ,-1);
  }			
  // Combine values of different files in one array.
  R_tot[(i*nfiles)+file][j]=R[i][j];
  R_tot_err[(i*nfiles)+file][j]=R_err[i][j];
}}
\end{verbatim}

\section{Stack design for the 55 MeV stack}\label{app:highstack}
\begin{table}[h!]
	\centering
	\caption[$55\,\rm MeV$ stack design. Kapton adhesive tape envelopes every thin foil, and additional to $25.4\,\upmu\rm m$ of kapton there is also $38.1\,\upmu\rm m$ of acrylic adhesive degrading the beam.]{$\mathit{55\,Me\!V}$ stack design. Kapton adhesive tape envelopes every thin foil, and additional to $\mathit{25.4\mu m}$ of kapton there is also $\mathit{38.1\mu m}$ of acrylic adhesive degrading the beam.}
	\begin{tabular}{cccc}
		\hline
		Foil & Material & \begin{tabular}{c}Apparent \bigstrut[t]\\ thickness ($\upmu\rm m$)\\ \end{tabular} & \begin{tabular}{c}Actual  \bigstrut[t]\\ areal density ($\nicefrac{\rm{mg}}{\rm{cm}^2}$) \end{tabular}\\
		\hline
		SS3 & Stainless steel & 125 & 100.48 \bigstrut[t] \\
		Fe 1 & Iron & 25 & 20.22 \\
		Ti 1 & Titanium & 25 & 11.09 \\
		Cu 1 & Copper & 25 & 22.40 \\
		A1 & Aluminum & 2240 & 598.68 \\
		Fe 2 & Iron & 25 & 19.91\\
		Ti 2 & Titanium & 25 & 10.94\\
		Cu 2 & Copper & 25 & 22.32 \\
		A2 & Aluminum & 2240 & 599.71 \\
		Fe 3 & Iron & 25 & 20.00\\
		Ti 3 & Titanium & 25 & 11.25 \\
		Cu 3 & Copper & 25 & 22.49 \\
		C1 & Aluminum & 970 & 261.48 \\
		Fe 4 & Iron & 25 & 19.93\\
		Ti 4 & Titanium & 25 & 10.91\\
		Cu 4 & Copper & 25 & 22.38 \\
		C2 & Aluminum & 970 & 261.65 \\
		Fe 5 & Iron & 25 & 20.02\\
		Ti 5 & Titanium & 25 & 10.99 \\
		Cu 5 & Copper & 25 & 22.35 \\
		C3 & Aluminum & 970 & 261.64 \\
		Fe 6 & Iron & 25 & 20.21\\
		Ti 6 & Titanium & 25 & 11.15 \\
		Cu 6 & Copper & 25 & 22.43 \\
		C4 & Aluminum & 970 & 261.71 \\
		Fe 7 & Iron & 25 & 19.93\\
		Ti 7 & Titanium & 25 & 11.17 \\
		Cu 7 & Copper & 25 & 22.24 \\
		H2 & Aluminum & 127 & $\sim\,$34.3 \\
		SS4 & Stainless steel & 125 & 101.25 \\
		\hline
	\end{tabular}
	\label{tab:highstack}
\end{table}
\clearpage
\newpage

\section{Investigating initial beam spread as cause for beam energy deviation}\label{app:spread}
At one point, the suspect of being responsible for the mismatch in beam energies, was the incident beam itself. More precise, the initial beam spread and its widening while traversing the stack. This investigation, using the initial data, before the adhesive was discovered as potential cause, leads to the conclusion that the effects from beam spread are small in comparison. On that note it might be valuable to state here, that the angular spread was eliminated as suspect very early in the experiment as well. Here the premise was, that the angular distribution would grow, while going through the stack.\par
\begin{figure}[h!]
	\centering
	\includegraphics[width=435px, keepaspectratio]{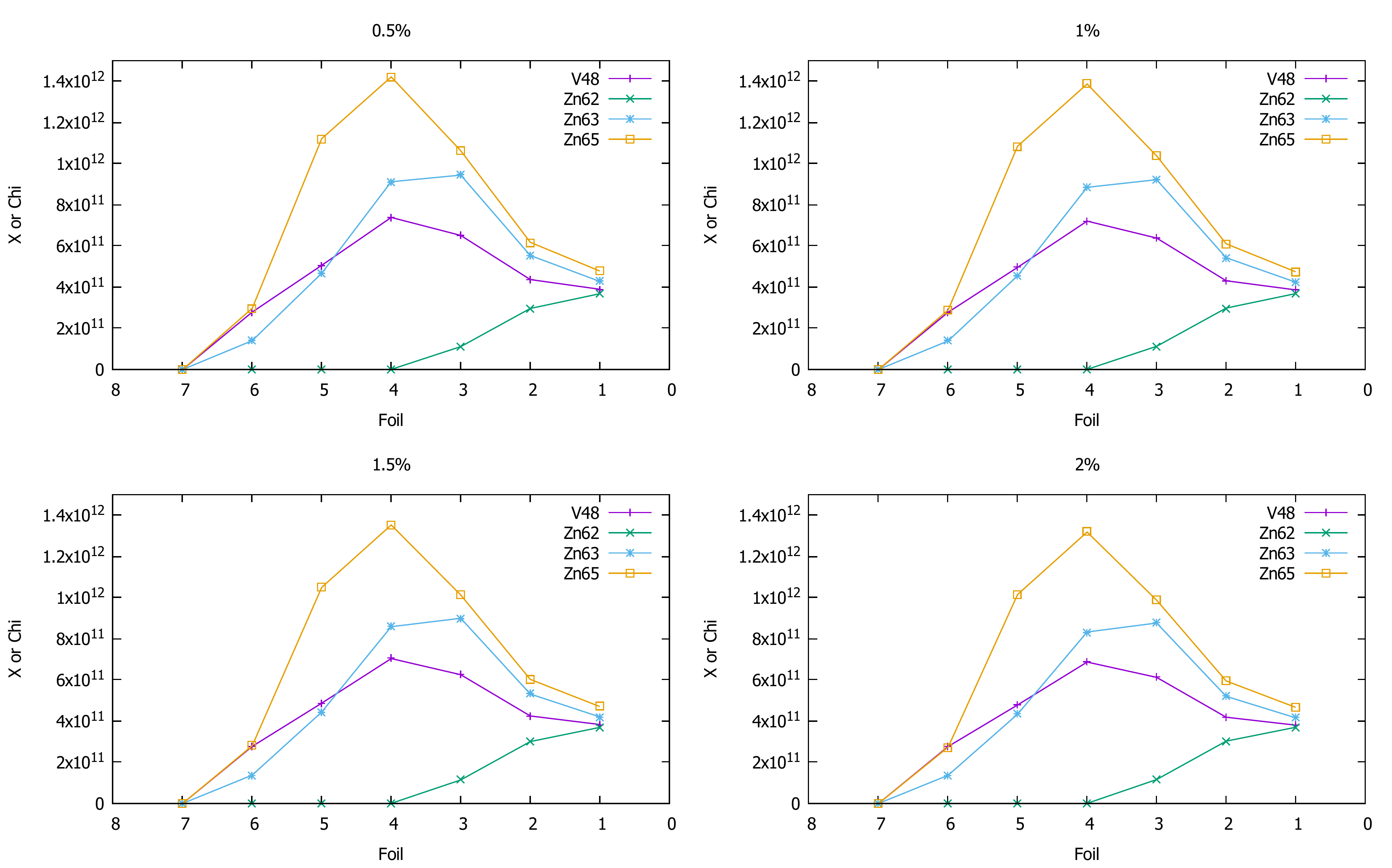}\\
	\caption{X values comparing the different monitor channels for each beam spread.}
	\label{fig:spread1}
\end{figure}

\begin{figure}[h!]
	\centering
	\includegraphics[width=435px, keepaspectratio]{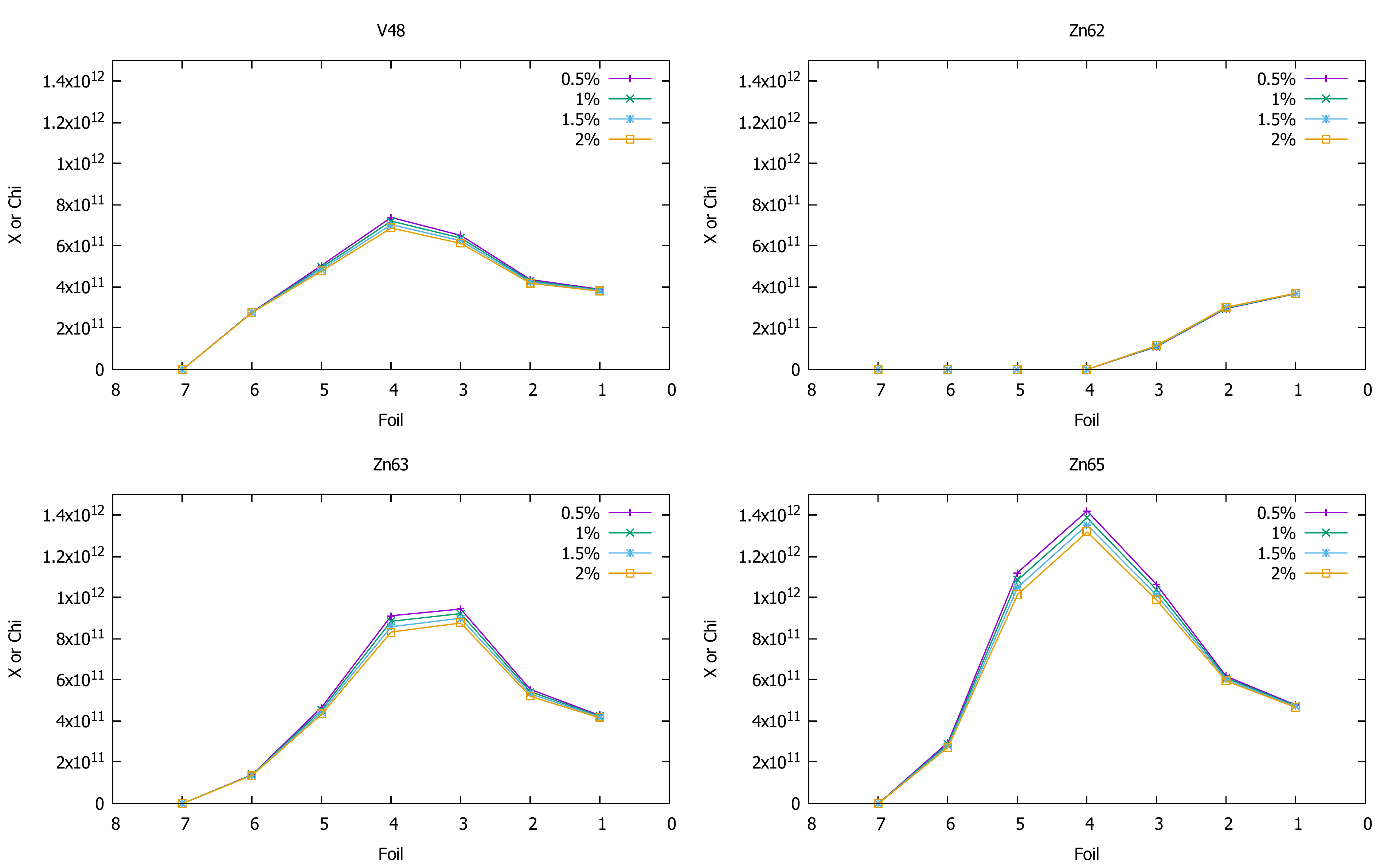}\\
	\caption{X values comparing the different beam spreads for each monitor channel.}
	\label{fig:spread2}
\end{figure}

\begin{figure}[h!]
	\centering
	\includegraphics[width=435px, keepaspectratio]{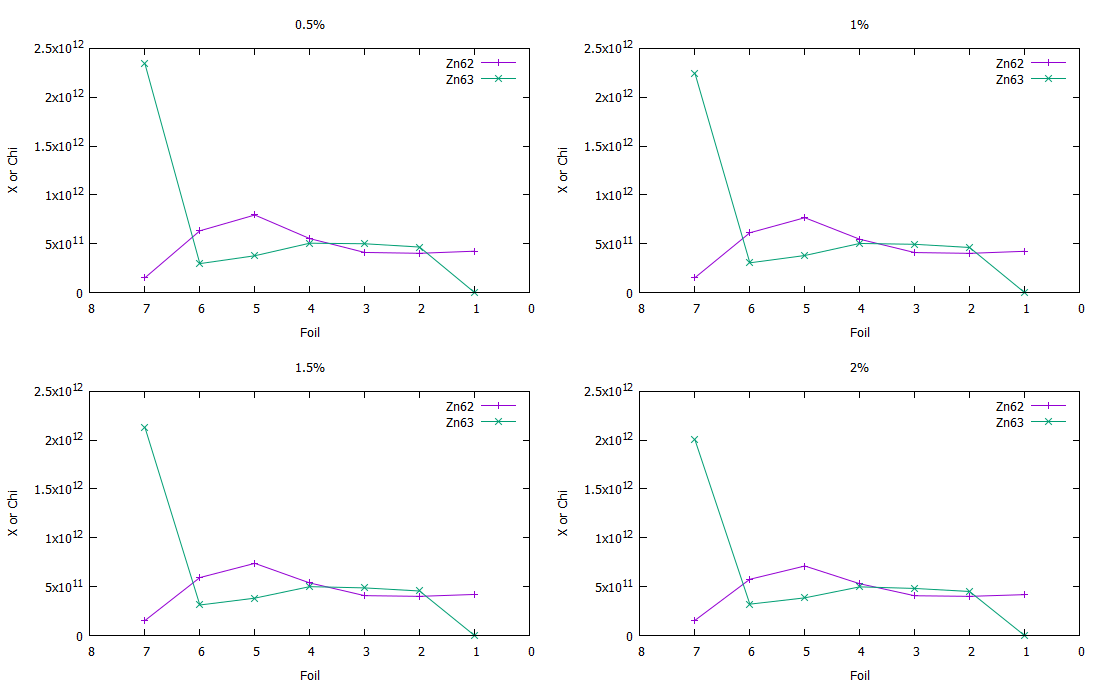}\\
	\caption{X values comparing the different monitor channels for each beam spread for the high energy stack.}
	\label{fig:spread3}
\end{figure}

\begin{figure}[h!]
	\centering
	\includegraphics[width=435px, keepaspectratio]{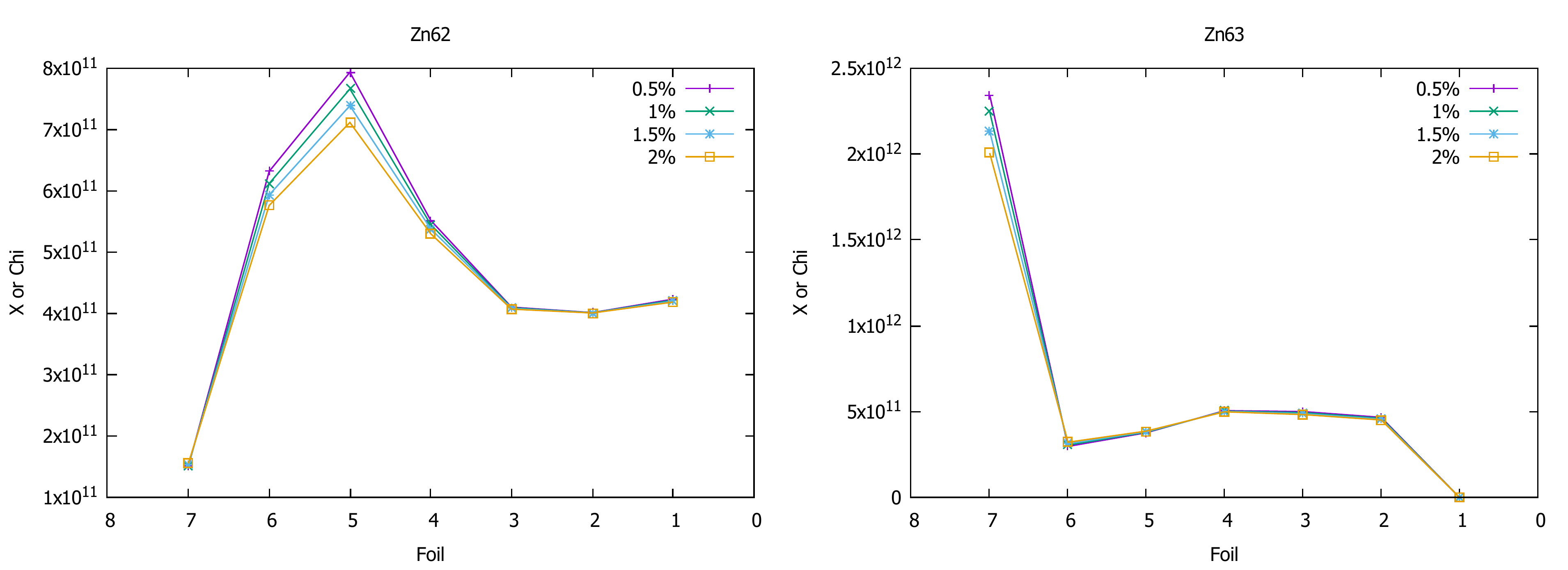}\\
	\caption{X values comparing the different beam spreads for each monitor channel for the high energy stack.}
	\label{fig:spread4}
\end{figure}

 
In the two Figures \ref{fig:spread1} and \ref{fig:spread2}, the X value is plotted over the foil position with the high energy, or incident side, to the right. All four monitor channels for the low energy stack are compared with each other and the change that different initial beam spreads ($0.5\,\%$, $1\,\%$, $1.5\,\%$ and $2\,\%$) would cause. It is clear that the difference between the monitors is much larger than between the different spreads. Meaning the root cause lies somewhere else. The fact that there are peaks in the graph, stems from the shape of the cross section. This comes about, because the energy is shifted in relation to the cross section peak.
\clearpage
\newpage
In Figures \ref{fig:spread3} and \ref{fig:spread4} the two monitor channels of the high energy stack are compared. The effect caused by defining different initial beam spreads in the simulation, is as small as for the low energy stack. 
\clearpage
\newpage

\section{Additional plot of the proton flux for the preliminary MCNP simulation}\label{app:flux-log}
\begin{figure}[hbt]
	\centering
	\includegraphics[width=420px, keepaspectratio]{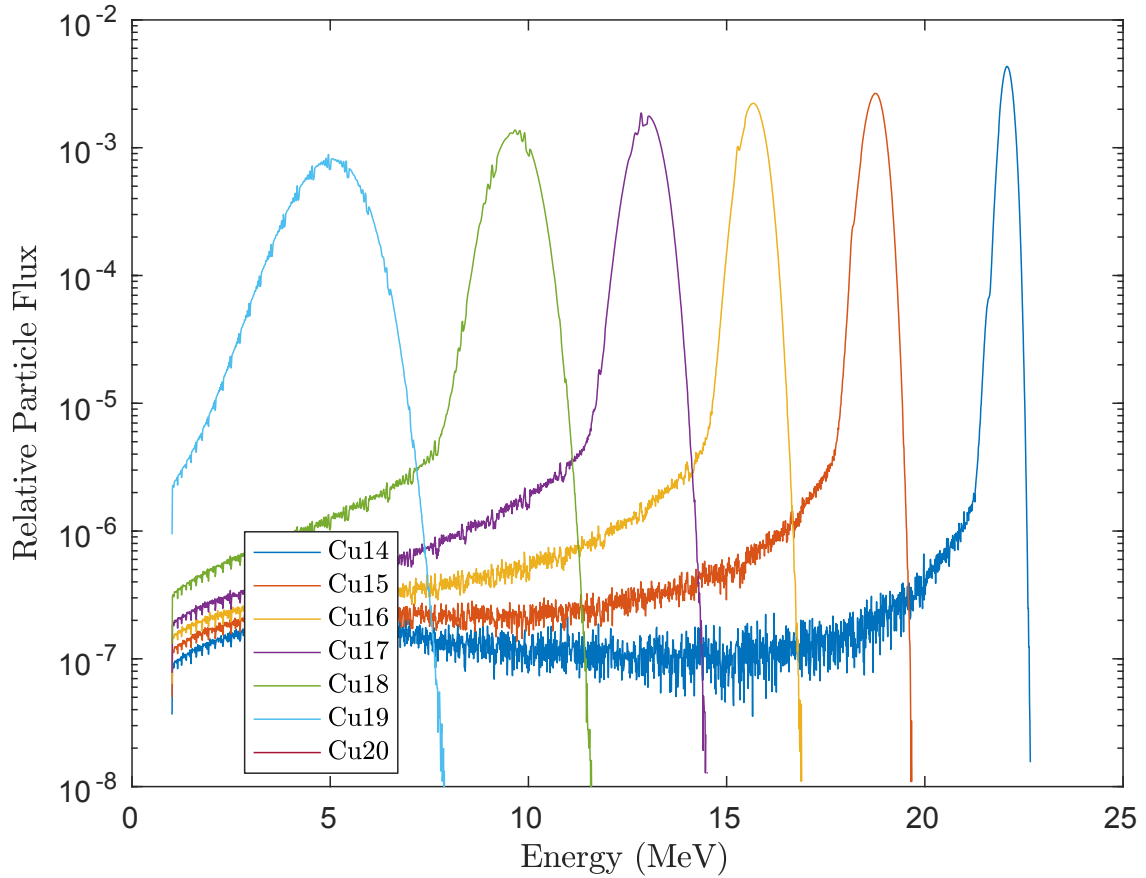}\\
	\caption{MCNP proton flux in the copper foils of the low energy stack. The flux actually only decreases slightly, but the energy distribution widens. Find the liner plot in Chapter \ref{chap:analysis}, Figure \ref{fig:flux}.}
	\label{fig:flux-log}
\end{figure}
\clearpage
\newpage

\section{Details on scaling factor for the preliminary MCNP model}\label{app:xlin}
\begin{figure}[hbt]
	\centering
	\includegraphics[width=420px, keepaspectratio]{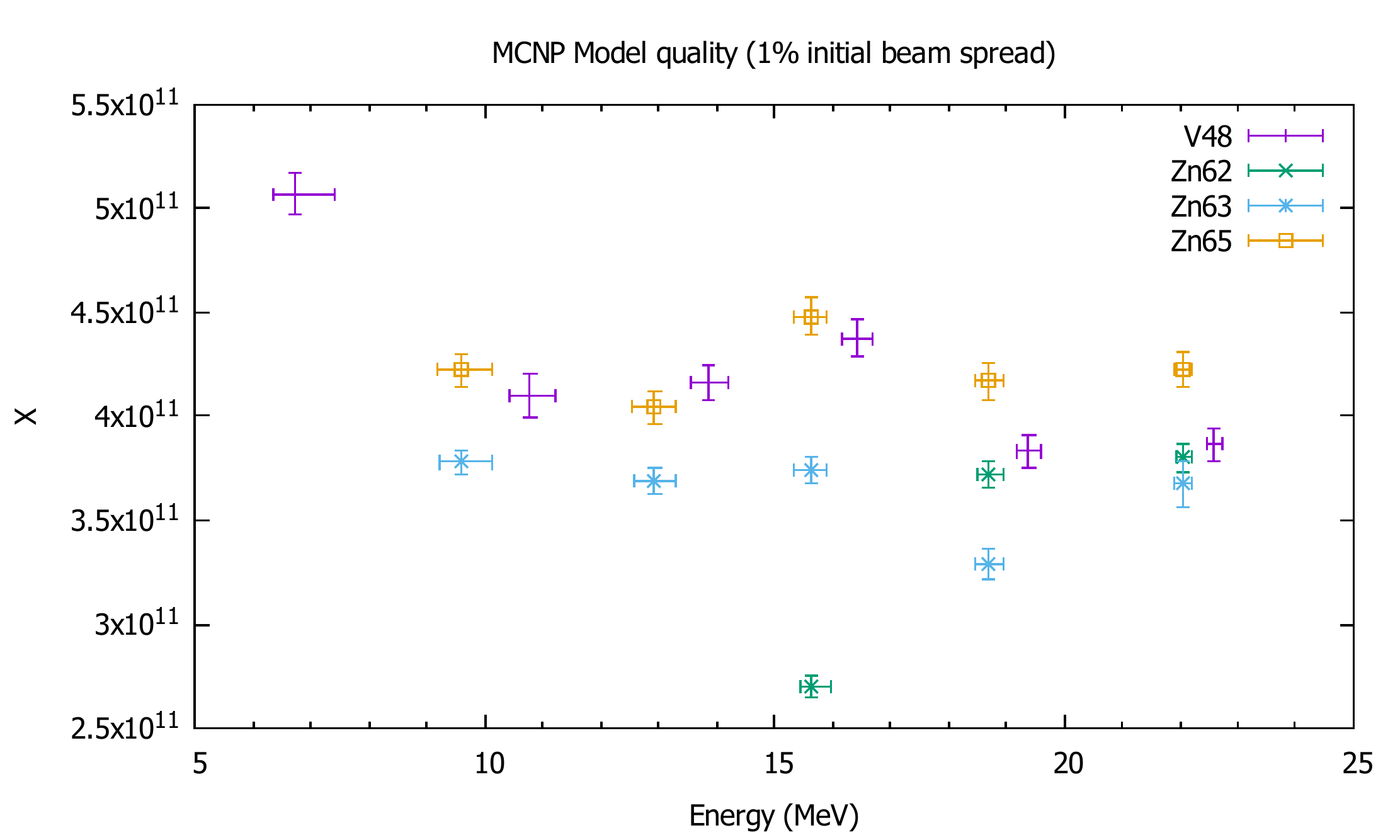}\\
	\caption{X values comparing the different monitor channels. Two outliers are cropped in this view, to better display the rest of the data. Find the complete plot, with the outliers, in Figure \ref{fig:x}}
	\label{fig:xlin}
\end{figure}
\clearpage
\newpage

\section{ROOT macro to evaluate MCNP model}\label{app:flux.c}
\begin{verbatim}
// ************************************************************ //
// Calculations
// ************************************************************ //
// Step 1: Finding the energy centroid and uncertainty of the
// monitors. This is achieved by calculating the flux weighted 
// average, becoming the centroid. Then we convolve CS and Flux and 
// search for one Sigma (34.1% above and below), resulting in 
// asymmetrical Energy uncertainties. For the most precise result.
Double_t E_centr[7];
// asymmetric error bars that are going to  be included in Step 2
Double_t E_uncert_low[7]; 
Double_t E_uncert_high[7];	
for(int j=0; j<7; j++){
 Double_t sum_conv = 0;
 Double_t sum_tot = 0;
 for(int i=0; i<bins; i++){
  sum_conv = sum_conv + (Flux_[j][i] * E_[i]);
  sum_tot = sum_tot + Flux_[j][i];
 }
 E_centr[j]=sum_conv/sum_tot;
 printf("Foil %g BeamEnergy centroid  = %g\n", j+1, E_centr[j]);
 EE_centr[mode*7+j]=E_centr[j];
}
// Step 2 a): using Cu and Ti foils with the IAEA cross sections. 
// Summing it up and create X, the correction factor we have to 
// multiply the model Rate by to match our experimental data. 
// X should be constant to be a perfect match.
// What we have:
// Array of the Energy E_[bin]
// Array of fluxes like Cu[7][bin] in Flux_[7][bin]
// Array of Rates R[7] and Nuclei/Area N[7]
Double_t X_[7];
Double_t X_uncert_low[7];
Double_t X_uncert_high[7];
Double_t F_[7][bins];
for(int j=0; j<7; j++){		
 Double_t sum = 0;
 Double_t X;
 Double_t X_uncert;
 for(int i=0; i<bins; i++){
  sum = sum + (Flux_[j][i] * CS_[i]*10e-27);
 }
 X=(R[j])/(sum*N[j]);
 X_uncert=sqrt( pow((R_err[j])/(sum*N[j]),2) 
              + pow(R[j]*N_err[j]/(sum*pow(N[j],2)),2) );
 // prevent division by zero: 
 if(sum==0){ 
  X=0;
  X_uncert=0;
 }
 X_uncert_low[j]=X-X_uncert;
 X_uncert_high[j]=X+X_uncert;
 X_[j]=X;
 printf("X[%g] = %g\n", j, X);
 printf("X_uncert = %g\n", X_uncert);
 XX_[mode*7+j]=X;
 XX_uncert_low[mode*7+j]=X_uncert_low[j];
 XX_uncert_high[mode*7+j]=X_uncert_high[j];
 
 // Step 2 b): Finding the asymmetric errorbars
 // Finding the asymmetric errorbars does not work if the Rate is 0 
 // or in an other way CS and Rate completely disagree.
 Double_t partsum = 0;
 bool low_set = false;
 bool set = false;
 bool high_set = false;
 for(int i=0; i<bins; i++){			
  partsum = partsum + (Flux_[j][i] * CS_[i]*10e-27);
  F_[j][i]= partsum/sum;
  if (F_[j][i]>=0.5-0.341 && !low_set){
   E_uncert_low[j] = E_[i];
   low_set=true;
  }
  if (F_[j][i]>=0.5 && !set){
   set=true;
  }
  if (F_[j][i]>=0.5+0.341 && !high_set){
   E_uncert_high[j] = E_[i];
   high_set=true;
  }
 }
 EE_uncert_low[mode*7+j] = E_uncert_low[j];
 EE_uncert_high[mode*7+j] = E_uncert_high[j];
}
\end{verbatim}

\chapter*{Acknowledgment\\ Danksagung}
\addcontentsline{toc}{chapter}{Acknowledgment}
\renewcommand{\chaptermark}[1]{\markboth{#1}{}}
\chaptermark{Acknowledgment}

There are a number of people without this work in particular would not have been possible for me to complete. I want to thank them for their support:\\
\begin{itemize}[label=$\circ$]
	\item Prof. Dr. Guido Drexlin for giving me the opportunity to complete my master's degree with an external master thesis under his supervision.
	
	\item Prof. Dr. Lee A. Bernstein for taking me into his team, teaching me and including me into the community at LBNL. I want to especially express my gratitude towards him for taking the time and comment in detail on the thesis work at every step of the way.

	\item Andrew Voyles for working together with me on the immediate challenges of the experiment and the analysis. His experience was invaluable. He displayed patience that I have rarely encountered, not just with me, but with anyone who happened to find themselves in need of his help.

	\item Haleema Zaneb for good teamwork on this and her work, on yttrium cross sections. Many of the challenges were mirrored and talking them over in a similar context was really helpful.

	\item All the other people at the LBNL 88-Inch Cyclotron and the Nuclear Data Group. Learning from you was my privilege.\par
\end{itemize}
\ \\
Besonderer Dank gilt meinen Eltern, Gabriele und Eduard Springer, die es mir erm\"oglichten ein Studium der Physik zu verfolgen. Sie haben mir immer unterst\"utzend zur Seite gestanden, bei all den Entscheidungen mit denen ich auf dem weg konfrontiert war.  \\
\ \\
Au\ss erdem m\"ochte ich ganz herzlich meinem Bruder Andreas und seiner Frau Kristin Springer zu der Geburt ihres Sohnes Sebastian gratulieren.

\end{document}